%% file: main.tex
\documentclass[twocolumn]{autart}

\usepackage[title]{appendix}
\usepackage{amsmath}
\usepackage{amsfonts}
\usepackage{amssymb}
\usepackage{extarrows}
\usepackage{psfrag}
\usepackage{graphicx}
\usepackage{color}
\usepackage{cite}
\usepackage{tabularx}
\usepackage{array}
\usepackage[dvipsnames]{xcolor}
\usepackage{soul}

\input{steve.tex}

\usepackage{pifont}
\usepackage{caption}
\usepackage{subcaption}
\usepackage{multirow}
\newcommand{\mycaption}[1]{\stepcounter{table}\raisebox{-7pt}
  {\footnotesize Table \thetable.\hspace{3pt} #1}}
 
\newcommand{\xmark}{\ding{55}}
\begin{document}
%\title{Simply Structured Distributed Estimators for Linear Systems}
%\maketitle
%\thispagestyle{empty}
%\pagestyle{empty}

\begin{frontmatter}
%\runtitle{Insert a suggested running title}  % Running title for regular 
                                              % papers but only if the title  
                                              % is over 5 words. Running title 
                                              % is not shown in output.

\title{{ Split-Spectrum Based Distributed State Estimation for Linear Systems}\thanksref{footnoteinfo}} % Title, preferably not more 
                                                % than 10 words.

\thanks[footnoteinfo]{This paper was not presented at any IFAC 
meeting. Portions of this paper were presented %, in abbreviated form and without proofs,
  at 2019 American Control Conference \cite{Lili19ACC} and IEEE Conference on Decision and Control \cite{Lili19CDC}. Corresponding author: Lili~Wang. 
%Tel. +XXXIX-VI-mmmxxi. 
%Fax +XXXIX-VI-mmmxxv.
}

\author[uci]{Lili Wang}\ead{lili.wang.zj@gmail.com},    % Add the 
\author[sb]{Ji Liu}\ead{ji.liu@stonybrook.edu},               % e-mail address 
\author[anu]{Brian D. O. Anderson}\ead{brian.anderson@anu.edu.au}
\author[yale]{A. Stephen Morse}\ead{as.morse@yale.edu},  % (ead) as shown

\address[uci]{Samueli School of Engineering,
University of California, Irvine}  % Please supply                                              
\address[sb]{Department of Electrical and Computer Engineering, Stony Brook University}        % here.

\address[yale]{Department of Electrical Engineering, Yale University}             % full addresses

\address[anu]{School of Engineering, Australian National University}
          
                             % keyword list or with the 
                                          % help of the Automatica 
                                          % keyword wizard

\begin{abstract}
This paper studies a distributed state estimation problem for both continuous- and discrete-time linear systems. A simply structured distributed estimator \{comprising interconnected local estimators\} is first described for estimating the state of a continuous and multi-channel linear system whose sensed outputs are distributed across a fixed multi-agent network. The estimator is then extended to non-stationary networks whose graphs switch according to a switching signal.
% with a fixed dwell time or  a variable but  fixed average dwell time, or switch arbitrarily under appropriate assumptions. 
The estimator is guaranteed to solve the problem, provided a network-widely shared high gain condition achieving a form of spectrum separation is satisfied. 
As an alternative to sharing a common gain across the network, a fully distributed version of the estimator is also studied in which each agent adaptively adjusts a local gain, though the practicality of this approach is subject to a robustness issue common to adaptive control. 
 A discrete-time version of the distributed state estimation problem is also studied, and a corresponding estimator based again on spectrum separation, but not high gain, is proposed for time-varying networks. For each scenario, it is explained how to construct the estimator so that the state estimation errors in the local estimators all converge to zero exponentially fast at a fixed but arbitrarily chosen rate, provided the network’s graph is strongly connected for all time. The proposed estimators are inherently resilient to abrupt changes in the number of agents and communication links in the inter-agent communication graph upon which the algorithms depend, provided the network is redundantly strongly connected and redundantly jointly observable. 
\end{abstract}

\end{frontmatter}

\section{Introduction}\label{sec:intro}
\vspace{-0.1in}

%\subsection{Background}
With the  growing interest in  sensor networks and
multi-agent systems,
the problem of estimating the state of a dynamical system whose measured
outputs are distributed across a network has been under  study in one form or another
 for a number of years
\cite{Khan2011,shamma,metro2,saber2,bullo.observe,sanfelice,martins,TAC.17,MitraPurdue2016,
Kim2016CDC,trent,trent2,CDC17.1, Lili19ACC}.

Depending on the nature of the system to be estimated, the distributed estimation problem has continuous- and discrete-time versions. 
In its simplest form, the continuous-time version of the distributed state estimation
 problem
 starts with
a network of $m>1$ agents
labeled $1,2,\ldots,m$ which are able
to receive information from their neighbors. Neighbor relations are characterized
by a directed graph $\mathbb{N}$, which may or may not depend on time,  whose
     vertices correspond to agents and whose arcs depict neighbor relations.
Each agent $i$  senses a
   signal $y_i\in\R^{s_i},\;i\in\mathbf{m} \dfb \{1,2,\ldots,m\}$ generated by a continuous-time system of the form
$\dot{x}=Ax,\;y_i(t) = C_ix, \; i\in \mathbf{m}$ 
and $x\in\R^n$. It is typically  assumed that $\mathbb{N}$ is strongly connected and
that the system  is jointly observable.
It is invariably assumed that each agent  receives certain real-time signals from its
 neighbors although what is received can vary from one problem formulation to the next.
 In all formulations, the goal is to devise local estimators, one for each agent, whose outputs
 are all asymptotically correct estimates of $x$.
 The local estimator dynamics for agent $i$ are typically assumed to depend only on the pair $(C_i,A)$
 and certain properties of $\mathbb{N}$.
The problem is basically the same in discrete time, except that rather than the
continuous-time model just described,  the
discrete-time model
$x(\tau+1)=Ax(\tau),\;y_i(\tau) = C_ix(\tau),\;i\in \mathbf{m},\; x\in\R^n$  is considered instead. More precise problem formulations will be given later. 

\vspace{-0.1in}

\subsection{Background}

\vspace{-0.1in}

 The study of distributed state estimation for linear systems can be dated back to the so-called distributed Kalman filter problem \cite{Olfati-Saber2005}, which involves system and measurement noise in the  problem formulation and has been widely studied for years \cite{Olfati-Saber2007,Khan2011}. 
Most available Kalman filter based approaches \cite{Khan2011,saber2,Olfati-Saber2005,Olfati-Saber2007,shamma} require the agents to both share ``signal information'', which  can be measurements or local state estimates, and fuse certain ``structural information'', which forms the covariance or information matrix of the nominal centralized Kalman filter for  each agent. 
 For the problem just described, the existing literature based on only ``signal information'' sharing can be classified into two categories, namely continuous- and discrete-time estimators/observers, except for our earlier work of \cite{CDC17.1,Automatica22} in which a hybrid observer was proposed for a continuous-time linear system.
 
 \vspace{-0.1in}

Continuous-time distributed  estimators have recently appeared in
\cite{Lili19ACC,Kim2016CDC,trent,TAC.17, trent2,ACC17.3}.
 By recasting   and then solving the distributed estimation problem as a 
 classical decentralized control
problem, the resulting estimator becomes  capable of
 estimating  the state  at a pre-assigned exponentially fast rate, assuming
  $\mathbb{N}$ is a constant strongly connected graph \cite{TAC.17}.
The work of \cite{Kim2016CDC}  seeks to
propose a distributed estimator for a
continuous-time system at the expense of certain design flexibility.
This is done, in essence, by  exploiting the $A$-invariance of  the
unobservable spaces of the pairs $(C_i,A)$; this in turn enables one
to ``split'' the local estimators into two parts, one based on
conventional spectrum assignment techniques  applied to the observable part of the state at each local estimator  and the other
based on consensus  among the unobserved parts of the state at each local estimator. The two parts are interacting but the use of a high gain serves to simplify the stability issue because of a split in the spectrum arising from   the design of the estimator.   The idea has been further developed in  \cite{trent, trent2,Kim20}. Specifically,   these latter references start to move beyond a restriction in  \cite{Kim2016CDC} permitting only constant, undirected, connected neighbor graphs to be addressed.
 The work of \cite{trent,
trent2} extends the result of \cite{Kim2016CDC} to the case when the
neighbor graph is constant, directed, strongly connected, while requiring that one   chooses gains to ensure   that certain
LMIs hold which are difficult to grasp intuitively. 
%{\color{red} Consider inserting some kind of interpretive comment, if true, like 'These may be difficult to solve'. or 'These are difficult to grasp intuitively'. or Conditions for solvablity are not provided/not transperent/not a problem, which is surprising'.(These are just random examples illustrating what I mean by an interpretive comment. Their cognitive content may be rubbish) }
In \cite{Lee2020}, motivated by a distributed least squares problem, a modified algorithm is proposed to deal with measurement noise constant, undirected, connected neighbor graphs.
%{\color{red} Some remark seems needed here about the nature of the noise. Sample remarks, but I do not know if they are true: The noise is assumed to be zero mean, white and gaussian, but its variance does not have to be known. OR Even though for the purposes of design, it must be assumed that the noise is zero mean, white and gaussian of known variance, the resulting estimators display attractive behavior even when the noise assumptions are not fulfilled. Or (desirably): {\bf{Something which allows us to say what we do is better}}. } 
 A distributed adaptive algorithm has recently been proposed in \cite{Kim20} which allows agents to join or leave the network over time, provided the resulting agent network always remains jointly connected and joint detectable.   An evident disadvantage of all these existing continuous-time distributed estimators is that they require a somewhat complicated gain computation procedure, and partially because of this, do not, at least not directly, admit discrete-time counterparts. 

\vspace{-0.1in}

 Discrete-time distributed   estimators have been recently studied in
\cite{Doos2013,Khan2014,Park2012,Park2012a,martins, MitraPurdue2016,  Mitra2019FiniteTimeDS, Acikmese2014,Ugrinovskii2013,Rego2021}.
Notable among them is the paper \cite{martins}.   Published prior to the appearance of the early continuous time paper \cite{TAC.17} applying to the same class of distributed systems, \cite{martins} solves the discrete-time 
distributed estimation problem for jointly observable, linear systems with constant, directed, strongly connected neighbor graphs. It builds on the idea of  recasting
the estimation problem as a classical decentralized
control problem.
Although   the  observer is  limited 
  to  discrete-time systems, it has been proved possible to  make use of the ideas in
   \cite{martins} to obtain,  as noted earlier, a distributed observer for
 continuous-time systems \cite{TAC.17}, but still for constant neighbor graphs. There are however other discrete-time distributed observers/estimators which do not admit continuous-time extensions,  illustrating that passage between discrete-time and continuous-time thinking may be harder than intuition initially suggests for distributed estimation problems.  
% The algorithm in \cite{Doos2013}
% works for fixed graphs with a relatively complicated topology design by studying the roles of each agent in the network.
% The distributed observer proposed in   \cite{Khan2014}  can track the system only if the so-called ``scalar tracking capacity'' condition is satisfied. 
By expanding on earlier work in \cite{MitraPurdue2016},
the papers \cite{Mitra2019FiniteTimeDS,MitraTAC2} provide a  procedure
 for constructing a centralized designed distributed observer for time-varying neighbor graphs. It requires the sharing of an index that records the age of the information across the network, and the agents are designed to act in a sequential manner to do state estimation.  The resulting algorithm, which is tailored exclusively to discrete-time systems,
requires a network-wide initialization step that serves to sort the agents in a specific order. Thereby it can deal with state estimation under assumptions that are weaker than
strong connectivity.

% {\color{cyan} Another aspect of the past discrete-time contributions is the extent to which the estimators} incorporate two time scales. An agent obtains data from its neighbors at a rate $q$ times that at which it obtains measurement data\cite{Khan2008,Acikmese2014,Khan2011,carli2}. {\color{red} try and stick in something like one of the following inserts or something similar. Write the insert to promote the sense of differentation between our paper and these earlier papers, but of course, do not misrepresent the other papers.}{\color{cyan} It is clear from a study of these papers that the use of two time scales flowed from a realization that stability issues could only be satisfactorily resolved by using the associated spectrum separaion. OR Neither paper makes the suggestion that resolution of the stability issues for the distributed estimator rely on a separation achieved using two time scales; rather, the two time scales appear to have been introduced ....give reason here. }

\begin{table*}[!t]
\centering
\mycaption{Comparison of Different Approaches to Design Distributed Estimators\label{tab:1}}
\begin{tabular}{ | m{3cm} | m{2cm}| m{2cm}  | m{2cm}| m{2cm} | m{2cm}| }\hline
Nature of Approach & Reference &Continuous-time Systems  & Discrete-time Systems & Exponential Convergence & Time-varying Graphs \\ 
\hline
\multirow{2}{3cm}{Kalman Filter Based Approach}&\cite{Khan2011,saber2,Olfati-Saber2005,Olfati-Saber2007}&   \xmark& \checkmark& \xmark& \xmark \\ \cline{2-6}
  & \cite{shamma}&  \checkmark& \xmark& \xmark& \xmark  
  \\ 
\hline
\multirow{2}{3cm}{Observability Decomposition Based Approach}&\cite{MitraPurdue2016}&   \xmark& \checkmark& \checkmark& \xmark \\ \cline{2-6}
  & \cite{Mitra2019FiniteTimeDS,MitraTAC2}&  \xmark& \checkmark& \checkmark& \checkmark  
  \\ 
 \hline
 \multirow{2}{3cm}{Decentralized Control Based Approach}& \cite{martins,Park2012,Park2012a} &  \xmark& \checkmark& \checkmark& \xmark 
 \\\cline{2-6}
&
 \cite{TAC.17,ACC17.3} &  \checkmark& \xmark& \checkmark& \xmark\\ 
 \hline
\multirow{3}{3cm}{Split-Spectrum Based Approach}&\cite{Lili19ACC,Kim2016CDC,trent,trent2}&   \checkmark& \xmark& \checkmark& \xmark \\ \cline{2-6}
  & \cite{Lili19CDC}&  \xmark& \checkmark& \checkmark& \xmark  
 \\\cline{2-6}  & This work&   \checkmark& \checkmark& \checkmark& \checkmark \\ \hline
\end{tabular}
\end{table*}
\vspace{-0.1in}

 Different approaches to the distributed state estimation problem are summarized in Table $1$.
 It turns out that the current paper is the first paper that can deal with both continuous-time and discrete-time systems while ensuring exponential convergence under time-varying neighbor graphs.
 
 The contribution of this paper rests on the following three  distinguishing features, differentiating it and highlighting it as a development of earlier work:

\vspace{-0.1in}

\begin{itemize}
    \item The paper describes   a simply structured, unified approach to the distributed state estimation problem and to design and analyze the corresponding distributed estimators for both continuous- and discrete-time linear systems with possibly time-varying graphs.   It is termed the ``split-spectrum''
approach because it ``splits'' the system spectrum into disjoint subsets corresponding to observable and unobservable subspaces.  In continuous time, this is achieved by a high gain mechanism, but in discrete-time by a different mechanism, viz. the adoption of two integrally related sampling rates. Though the mechanisms are instrumentally different, their purpose is fundamentally the same. 
%Several well-known properties of invariant subspaces are exploited in the analysis.  
  It is termed `unified' because the approach is shown to work for both continuous- and discrete-time linear systems over both constant and time-varying neighbor graphs.
\item A fully distributed version of the estimator is   separately  studied where each agent can adaptively adjust a local gain, with simpler gain computation procedure and analysis compared with \cite{Kim20}.
\item  Exponential convergence of the error dynamics is ensured with an arbitrarily assigned convergence rate. A great advantage of our methodology is that the designs and algorithms developed under a noiseless assumption are then necessarily tolerant of some level of noise, simply because we take care to ensure an exponential convergence.
\end{itemize}
 \vspace{-0.1in}

% All the proposed continuous- or discrete-time distributed estimators, except for the adaptive one, are inherently resilient to abrupt changes in the number of agents and communication links, provided the network is redundantly strongly connected and redundantly jointly observable, with carefully design before the algorithms start.
% Formal definitions of the redundant strong connectivity and redundant joint observability concepts (which are fairly intuitive) can be found in \cite{Automatica22}.
\vspace{-0.05in}
 
 It is assumed in this paper that the neighbor graph of the network is always strongly connected.  From the perspective of the real world,  requiring the underlying network to be strongly connected ``at every time step'' is an assumption that will occur in a great many (though obviously not all) cases, and as such, is deserving of a separate study in its own right.
The extension to more general time-varying graphs is one future direction.  
It may not be conceptually difficult, however intricate the details may be.

\vspace{-0.1in}

The paper first describes the split-spectrum based distributed estimator for the case when the system dynamics are continuous with a stationary network in \S \ref{sec:split_continuous_estimator}, and with associated background analysis is given in \S \ref{sec:continuous_constant_graph}. The estimator is then extended to deal with non-stationary networks whose neighbor graphs switch according to a switching signal with a fixed dwell time or a variable dwell time with prescribed average, the ideas being detailed in   \S \ref{sec:continuous_avg_dwell}. In the case when the interconnection among the agents can always be  modelled usingdoubly stochastic matrices \{e.g., undirected graphs with the Metropolis weights \cite{metro2}\}, it is shown in \S \ref{sec:continuous_avg_dwell} that the estimator functions correctly even if the neighbor graph switches arbitrarily, provided the graph is always strongly connected. The estimators mentioned above all rely on the existence of  a sufficiently large, network-widely shared gain. A fully distributed version of the
estimator is then studied in \S \ref{sec:adaptive_gain} where each agent can adaptively adjust a local gain. The adaptive estimator is subject to a robustness issue common
to adaptive control. The proposed estimators, except for the adaptive one, are inherently resilient
to abrupt changes in the number of agents and communication links in the inter-agent communication graph upon which the
algorithms depend, an issue  which is discussed in \S \ref{sec:resilience}. 
Then the split-spectrum based estimator design is extended to the case when the system dynamics is discrete in \S \ref{sec:split_discrete} for both constant and time-varying neighbor graphs. 
Simulation validation is provided in \S \ref{sec:simu}.

%This paper extends the results in \cite{Lili19ACC} to time-varying neighbor graphs for a continuous-time system, and also provides the results for discrete-time system which was partially presented in \cite{Lili19CDC}. Moreover, both discussion and simulation results imply that  our distributed estimators are inherently resilient to connection failures.  

%\subsection{Organization}

\vspace{-0.1in}

The material in this paper was partially presented in \cite{Lili19ACC, Lili19CDC}, but this paper presents a more comprehensive
treatment of the work. Specifically, the paper crafts continuous-time distributed estimators for two types of non-stationary networks in \S \ref{sec:continuous_avg_dwell} and a fully distributed adaptive estimator in \S \ref{sec:adaptive_gain}, 
which were not included in \cite{Lili19ACC, Lili19CDC}.

\vspace{-0.2in}

\section{Continuous-Time Distributed Estimator}\label{sec:split_continuous}

\vspace{-0.1in}

We are interested in a network of $m>0$ \{possibly mobile\} autonomous agents labeled $1,2,\ldots, m$ which are able to  receive information from  their ``neighbors'', where by a {\em neighbor} of agent  $i$ is meant any other agent within agent $i$'s reception range. 
We write $\mathcal{N}_i(t)$ for the labels of agent $i$'s neighbors at time $t\in[0,\infty)$ and always take agent $i$ to be a neighbor of itself.
Neighbor relations at time $t$ are characterized by a directed graph $\mathbb{N}(t)$  with $m$ vertices and a set of arcs defined so that there is an arc in $\mathbb{N}(t)$ from vertex $j$ to vertex $i$  whenever agent $j$ is a neighbor of agent $i$ at time $t$.
Since each agent $i$ is always a neighbor of itself, $\mathbb{N}(t)$ has a self-arc at each of its vertices.
Each agent $i$ can sense a continuous-time signal $y_i\in\R^{s_i},\;i\in\mathbf{m}\dfb\{1,2,\ldots, m\}$, where
\begin{eqnarray}y_i &=&C_ix,\;\;\;i\in
  \mathbf{m}\label{eq:pre_sys1}\\\dot{x} &= &Ax\label{eq:pre_sys2}
\end{eqnarray}
and $x\in\R^n$.
We assume throughout that $C_i \neq 0,\;i\in\mathbf{m}$, and that the system defined by   \rep{eq:pre_sys1} and \rep{eq:pre_sys2} is \textit{ jointly observable}; i.e.,  with $C = \begin{bmatrix}C_1' & C_2' &\cdots &C_m'\end{bmatrix}'$, the matrix pair $(C,A)$ is observable.
% Joint observability is equivalent to the requirement that
% $$\bigcap_{i\in\mathbf{m}}\mathcal{V}_i = 0$$
% where $\mathcal{V}_i$ is the {\em unobservable space} of $(C_i,A)$; i.e. $\mathcal{V}_i = \ker \begin{bmatrix}C_i' &(C_iA)' & \cdots &(C_iA^{n-1})'\end{bmatrix}'$. 
% As is well known, $\mathcal{V}_i$ is the largest $A$-invariant subspace contained in the kernel of $C_i$.
Joint observability is
equivalent to the requirement that
$ \bigcap_{i\in\mathbf{m}}\mathcal{V}_i = 0$,
where $\mathcal{V}_i$ is the {\em unobservable space} of $(C_i,A)$; i.e.
$\mathcal{V}_i = \ker \begin{bmatrix}C_i' &(C_iA)' & \cdots &(C_iA^{n-1})'\end{bmatrix}'$. 
As is well known, $\mathcal{V}_i$ is
 the largest $A$-invariant subspace contained in the
kernel of $C_i$.
Generalizing the results that follow to the case when $(C,A)$ is only detectable is quite straightforward and can be accomplished using well-known ideas.
However, the commonly made assumption that each pair $(C_i,A),\; i\in \mathbf{m}$, is observable, or even just detectable, is very restrictive, grossly simplifies the problem and is unnecessary.
%It is precisely the exclusion of this assumption that distinguishes the problem posed here from almost all of the distributed estimator problems addressed in the literature. ({\color{red} literature})
The assumption $C_i\neq 0$ is not necessary provided the more relaxed problem is properly formulated.
The assumption is made for the sake of simplicity.
The problem of interest is to construct a suitably defined family of linear estimators in such a way so that no matter what the estimators’ initial states are,  each agent obtains an asymptotically correct estimate $x_i$ of $x$ in the sense that the estimation error $x_i(t)-x(t)$ converges to zero as fast as $\text{exp}(-\lambda t)$ does, where $\lambda$ is an arbitrarily chosen but fixed positive number.

\vspace{-0.1in}

%\section{Split-spectrum Based Estimator in Continuous Time}\label{sec:split_continuous}

This section proposes the estimator first, and then analyses the estimator beginning with the condition that  the neighbor graph $\mathbb{N}(t)$ is constant. A time-varying   neighbor graph $\mathbb{N}(t)$ is then considered in which changes occur  according to a switching signal. Later, a fully distributed algorithm based on use of multiplicative adaptive gain control is developed.
\vspace{-0.1in}

 \subsection{The Estimator} 
 \label{sec:split_continuous_estimator}
 \vspace{-0.1in}

% In the sequel it is assumed that the neighbor graph $\mathbb N$ is constant and strongly connected. Let $\mathcal{N}_i$ be the set of labels of agent $i$'s neighbors.

The estimator to be considered consists of $m$   local or private estimators  of the form for each $i\in\mathbf{m}$,
 \vspace{-0.1in}
\begin{equation}
    \dot{x}_i = (A+K_iC_i)x_i -K_iy_i  - gP_i\bigg (\!\!x_i-\frac{1}{m_i(t)}\!\!\!\sum_{j\in\mathcal{N}_i(t)
} \!\!\!x_j \!\!\bigg)\label{eq:split_continuous_observer}   
\end{equation}
where $m_i(t)$ is the number of labels in $\mathcal{N}_i(t)$, $g$ is a suitably defined positive gain  common to all local estimators, each $K_i$ is a suitably defined matrix, and for each $i\in\mathbf{m}$,  $P_i$ is the orthogonal projection on the unobservable space of $(C_i,A)$.
  The term $(A+K_iC_i)x_i-K_iy_i$ is designed to enable each agent $i$ to be able to recover the observable   part of the state by itself, and the term $-gP_i (x_i-\frac{1}{m_i(t)}\sum_{j\in\mathcal{N}_i(t)
} x_j )$  is for the purpose of utilizing information from its neighbors to recover the unobservable  part of the state. It will be shown that with this design, the spectrum of the system matrix can be split into two subsets.  Details on how to design the parameters $K_i$ and $g$ will be provided in the following.

 \vspace{-0.1in}

 \subsubsection{The Error Dynamics}\label{sec:error_continuous}
 \vspace{-0.1in}

 This subsection provides the error dynamics of the proposed estimator. It will be shown that the spectrum of the error system matrix can be split into two subsets based on the observability of each agent with $K_i$ defined properly. 
 
  First note from \eqref{eq:split_continuous_observer}, that the state estimation error $e_i=x_i-x$ satisfies
  \vspace{-0.1in}
\eq{\dot{e}_i = (A+K_iC_i)e_i - gP_i\left (e_i-\frac{1}{m_i(t)}\sum_{j\in\mathcal{N}_i(t)}e_j\right  )\label{eq:split_continuous_err}}

\vspace{-0.2in}

Consequently the overall error vector
   $e = \begin{bmatrix}
   e_1' \ \ldots\ e_m'
   \end{bmatrix}'$
satisfies
\eq{\dot{e} = (\bar{A} -gP(I_{mn}-\bar S(t)))e\label{eq:split_continuous_em}}
where 
 $\bar{A} = $ block diag $\{A+K_1C_1 ,A+K_2C_2,\ldots,A+K_mC_m\}$,
 $P = $ block diag $\{P_1 ,P_2,\ldots, P_m\}$,  $\bar S (t) =S(t)\otimes I_n $ with
$S(t) = D_{\mathbb{N}(t)}^{-1}\mathcal A'_{\mathbb{N}(t)}$. 
Here  $I_k$ is the $k\times k$ identity matrix, and $\mathcal A_{\mathbb{N}(t)}$ is the adjacency matrix of $\mathbb{N}(t)$ and
 $ D_{\mathbb{N}(t)}$
 is the diagonal matrix whose $i$th diagonal entry is the in-degree
  of $\mathbb{N}(t)$'s $i$th vertex. 
  Note that $\mathbb{N}(t )$ is the graph\footnote{The {\em graph} of
   an $n\times n$ matrix $M$ is  that directed graph
  on $n$ vertices possessing
 a directed arc from vertex  $i$ to vertex $j$ if $m_{ij}\neq 0$ \{p. 357, \cite{horn}\}.}
 of $S'(t)$ and that the diagonal entries of $S'(t)$
 are all positive because each agent is a neighbor of itself.
 The matrix $S(t)$ is evidently a stochastic matrix. 

 \vspace{-0.1in}

\begin{proposition}\label{prop:error}
The spectrum of the error system matrix $\bar A -gP(I_{mn}-\bar S)$ splits into two subsets.  One subset contains  the union of certain subsets of the eigenvalues associated with the $i$-th local estimator, $i\in\mathbf m$, these being  able to  be arbitrarily positioned by choice of the $K_i$ and independent of $g$. The second subset is  independent of the choice of the $K_i$, and depends, though not to the extent of being able to be arbitrarily positioned, on $g$.
\end{proposition}

 \vspace{-0.1in}
 
  We remark that in the proof below, we will explain how to choose the $K_i$ to ensure that the associated set of eigenvalues has degree of stability $\lambda$ (ensuring an estimation error decay at least as fast as $\exp(-\lambda t)$), while subsequently we will explain how to choose $g$ to ensure stability of the remaining part of the spectrum with the same minimum exponential decay rate.
 
 \vspace{-0.1in}
 
 { \bf Proof of Proposition \ref{prop:error}:}
The definitions of $K_i$ and $g$ begin with the specification of a desired convergence rate bound $\lambda>0$. 
To begin with, each matrix  $K_i$ is defined as follows.
For each fixed $i\in\mathbf{m}$,  write $Q_i$ for any full  rank matrix whose kernel is the unobservable space of $(C_i,A)$ and let $\bar{C}_i$ and $\bar{A}_i$ be the unique solutions to $\bar{C}_iQ_i = C_i$ and $Q_iA=\bar{A}_iQ_i$  respectively.  
Then  the matrix pair $(\bar{C}_i,\bar{A}_i)$ is observable.  
A matrix $\bar{K}_i$ can be chosen to ensure that  the convergence of $\text{exp}\{(\bar{A}_i + \bar{K}_i\bar{C}_i)t\}$ to zero is as fast as the convergence of $\text{exp}(- \lambda t)$ to zero is.
There are several well-documented ways to do this \{e.g,  spectrum assignment algorithms or Riccati equation solvers\},
since each pair $(\bar{C}_i, \bar{A}_i)$ is observable. Having chosen such $\bar{K}_i$, $K_i$
is then chosen to be $K_i = Q_i^{-1}\bar{K}_i$ where $Q_i^{-1}$ is a right inverse for $Q_i$.
The definition implies that  
 \vspace{-0.1in}
\begin{equation}\label{eq:QA1}Q_i(A+K_iC_i) = (\bar{A}_i+\bar{K}_i\bar{C}_i)Q_i\end{equation} and that
$(A+K_iC_i)\mathcal{V}_i\subset \mathcal{V}_i$. 
The latter, in turn, implies that
there is a unique matrix  $A_i$ which satisfies 
 \vspace{-0.05in}
 \begin{equation}
    \label{eq:QA2}
(A+K_iC_i)V_i = V_iA_i\end{equation} where $V_i$ is a basis
matrix\footnote{For simplicity, we assume that the columns of $V_i$ constitute an orthonormal basis for $\mathcal{V}_i$ in which case  $P_i = V_iV_i'$.} for $\mathcal{V}_i$.
%Prior to explaining how to choose $g$, it will be helpful to explain what
 %defining the $K_i$ in this way accomplishes.

Next we show what defining the $K_i$ in this way accomplishes.
Note that  the
subspace $\mathcal{V} = \mathcal{V}_1\oplus\mathcal{V}_2\oplus \cdots \oplus
\mathcal{V}_m$
 is
$\bar{A}$ - invariant  because $(A+K_iC_i)\mathcal{V}_i\subset
\mathcal{V}_i,\;i\in\mathbf{m}$. Next,  let $Q = $ block diag $
\{Q_1, Q_2,\ldots ,Q_m\}$ and $V = $ block diag $ \{V_1,
V_2,\ldots ,V_m\}$ where $V_i$ is a matrix whose columns form an orthonormal basis for $\mathcal{V}_i$. Then
 $Q$ is a full rank matrix whose
kernel is $\mathcal{V}$  and $V$ is a basis matrix for $\mathcal{V}$ whose
 columns form an orthonormal set.
 It follows that $P = VV'$,  that $QP=0$, and that
\begin{eqnarray}
Q\bar{A}& = &\bar{A}_VQ\label{eq:split_QA}\\
 \bar{A}V & = & V\tilde{A}\label{eq:split_AV}
 \end{eqnarray}
  where $\tilde{A} =\text{block diag}\;\{A_1,A_2,\ldots ,A_m\}$ and 
\begin{equation}
\label{eq:split_discrete_sunday}\bar{A}_V =  \text{block diag}\;
\{\bar{A}_1+\bar{K}_1\bar{C}_1 ,\ldots,
\bar{A}_m+\bar{K}_m\bar{C}_m\}.\end{equation}
 
%and $A_i$  is the unique solution to  $(A+K_iC_i)V_i=V_iA_i$. 
Let $V^{-1}$ be any left inverse for $V$ and let $Q^{-1}$ be
 that right inverse for $Q$ for which $V^{-1}Q^{-1} = 0 $. Then
 \vspace{-0.1in}
\begin{equation}
   \label{eq:split_continuous_split}
\bar{A} -gP(I_{mn}-\bar S(t))= T\begin{bmatrix}\bar{A}_V & 0\\    \hat{A}_V(t) &A_V(t)\end{bmatrix}T^{-1}\end{equation} 
where
$\hat{A}_V(t) = V^{-1}(\bar{A} -g(I_{mn}-\bar S(t)))Q^{-1}$ and
 $A_V(t) =\tilde{A}-gV'(I_{mn}-\bar S(t))V.$
Here $T = \begin{bmatrix}Q^{-1} & V\end{bmatrix}$. It is easy to check that
$T^{-1} =\begin{bmatrix}Q' & V\end{bmatrix}'$.
 
 \vspace{-0.1in}
 
 According to \eqref{eq:split_continuous_split}, the spectrum of  $\bar{A} -gP(I_{mn}-\bar S)$ is equivalent to the union of the spectrum of $\bar A_V$ and $A_V$.
\hfill $\qed$
 
 \vspace{-0.1in}
 
Recall that  the $\bar{K}_i$ have been already  been chosen so that each matrix
exponential $\text{exp}\{(\bar{A}_i+\bar{K}_i\bar{C}_i)t\}$
 converges to zero as fast
 as $\text{exp}({- \lambda t})$ does.
 Because of this and the fact that $\hat{A}_V(t)$ is a bounded matrix,
  to ensure that for each fixed $\tau$,
 the state transition matrix  $\Phi(t,\tau)$  converges to zero   as fast
 as $\text{exp}({-\lambda t})$ does, it is enough to  choose $g$ so that the state transition matrix of $A_V(t)$ converges
 to zero  at least as fast as $\text{exp}({-\lambda t})$ does.
The requisite  condition on $g$ is provided below for three different neighbor graph connectivity assumptions.

 \vspace{-0.1in}
\subsection{Constant Neighbor Graph}\label{sec:continuous_constant_graph}
 
 \vspace{-0.1in}
 
 This subsection focuses on the case when the neighbor graph $\mathbb{N}(t)$ is a constant graph $\mathbb{N}$.
 The following result can be obtained.
   
 \vspace{-0.1in}
\begin{theorem}\label{thm:split_continuous_constant}
 For any given positive number $\lambda$, if the neighbor graph  $\mathbb{N}$ is fixed and strongly connected, and the system defined by \eqref{eq:pre_sys1} and \eqref{eq:pre_sys2} is jointly observable, there are matrices $K_i$, $i\in \mathbf{m}$ such that for $g$ sufficiently large,
 each estimation error $x_i(t)-x(t)$ of the distributed estimator defined by \eqref{eq:split_continuous_observer}, converges to zero as $t\rightarrow \infty$ as fast as $\text{exp}({-\lambda t})$ converges to zero.
 \end{theorem}
  
 \vspace{-0.1in}
% The  theorem  proves to be a simple consequence of the following proposition.

 The proof of the theorem   involves making use of the following result, with all proofs being contained in Appendix A.  In particular, the proof of Proposition \ref{prop:split_continuous_stable} makes use of the properties of strong connectivity of the neighbor graph and joint observability of the system.
  
 \vspace{-0.1in}
 \begin{proposition}\label{prop:split_continuous_stable}
$-V'(I_{mn}-\bar S)V$ is a continuous-time stability matrix.
\end{proposition}

 \vspace{-0.1in}
\subsection{   Switching Neighbor Graph}\label{sec:continuous_avg_dwell}
 
 \vspace{-0.1in}
 
In the sequel the problem  is studied under the assumption that $\mathbb{N}(t)$ changes according to a switching signal with a fixed dwell time or
a variable  dwell time  with fixed average.
 To characterize the assumed
 time dependence of $\mathbb{N}(t)$,
let $\mathcal{G}=\{\mathbb{G}_1,\mathbb{G}_2,\ldots, \mathbb{G}_{n_g}\} $ denote the set of all directed,
strongly connected graphs on $m$ vertices which
 have self-arcs at all vertices; here $n_g$ is the number of graphs in $\mathcal{G}$.
In some situations, the switching signals always  have consecutive discontinuities separated by  a value which is no less than  a fixed positive real number  $\tau_D$. 
It is called a {\em dwell time}   \cite{asm:ecc.sup}.
In certain situations, the switching signals may occasionally have consecutive discontinuities separated by less than $\tau_D$, but for which the average interval between consecutive discontinuities is no less than $\tau_D$. This leads to the concept of an average dwell-time.
With $\tau_D$ and $\delta $ fixed define $\mathcal{S}_{\text{avg}} $ for the set of all piecewise-constant switching signals $\sigma:[0,\infty)\rightarrow \{1,2,\ldots, |\mathcal{G}|\}$ satisfying 
$\delta_{\sigma}(t_0,t)\leq \delta_0+\frac{t-t_0}{\tau_D}$.
Here $\delta_{\sigma}(t_0,t)$ denotes the number of discontinuities of $\sigma$ in the open interval $(t_0, t)$. 
The constant $\tau_D$ is called  the average dwell-time and
$\delta_0$  the chatter bound 
\cite{average}.
 By the set of all time-varying  neighbor
    graphs with average dwell-time $\tau_D$
    is meant the set
     $\{\mathbb{G}_{\sigma}:\sigma\in\mathcal{S}_{\text{avg}}\}$.
     Note that switching according to a dwell time  is a special case of switching according to an average dwell time. In the following,
it is assumed that $\mathbb{N}\in  \{\mathbb{G}_{\sigma}:\sigma\in\mathcal{S}_{\text{avg}}\}$.

The  problem  to which this subsection is addressed is this. For fixed averaged dwell-time  $\tau_D$ and the chatter bound $\delta_0$, devise a procedure
 for crafting $m$ local estimators, one for each agent, so that  for each neighbor graph
  $\mathbb{N}\in  \{\mathbb{G}_{\sigma}:\sigma\in\mathcal{S}_{\text{avg}}\}$,
 all $m$ state estimation errors converge to zero exponentially fast at a prescribed rate.

The estimator to be considered 
is still the same as the estimator described in \eqref{eq:split_continuous_observer}, with the exception that  $g$ is
 chosen differently. 
 %According to \S \ref{sec:split_continuous_estimator}, it remains to be shown that with $g$ sufficiently large, the state transition matrix of $A_V(t)$ converges to zero as fast as $\text{exp}({-\lambda t}) $ does. 
 %To accomplish this, use will be made of the following result.
The following result can be derived.

 \vspace{-0.1in}
 \begin{theorem}\label{thm:split_continuous_average}
For any fixed positive numbers $\tau_D$ and $\lambda$, there exists a positive number $g^*$ with the following property.
 For any  value of $g\geq g^*$, any neighbor graph $\mathbb{N}\in\{\mathbb{G}_{\sigma}:\sigma\in\mathcal{S}_{\text{avg}}\}$, if the system defined by \eqref{eq:pre_sys1} and \eqref{eq:pre_sys2} is jointly observable, there are matrices $K_i,\; i\in \mathbf m$ such that, each state estimation error
 $e_i = x_i-x,\;i\in\mathbf{m}$   of the distributed estimator defined by $\eqref{eq:split_continuous_observer}$ converges to zero as $t\rightarrow \infty$ as fast as $\text{exp}({-\lambda t})$ does.\end{theorem}
 
 \vspace{-0.1in}
 
 The proof of  Theorem \ref{thm:split_continuous_average} depends on the following lemma.
 
 \vspace{-0.1in}
 
 \begin{lemma}\label{lem:split_continuous_average}
Let $M_1,M_2,\ldots,M_{|\mathcal{G}|}$ be a set of   $n\times n$ exponentially stable real matrices associated with a set $\mathcal G=\{\mathbb G_1,\mathbb G_2,\dots,\mathbb G_{|\mathcal G|}\}$ of directed strongly connected graphs with self-arcs at all vertices. Let $\sigma$ denote the switching signal with average dwell time $\tau_{D}$ governing the selection of a graph from $\mathcal G$. 
Then for any $n\times n$ real matrix  $N$ and  positive number $\lambda$
 there is a positive
number $g^*$, depending on $\tau_D$ for which, for each $\sigma\in\mathcal{S}_{\text{avg}}$ and $g\geq g^*$,  all solutions  to
\eq{
\dot{x} =(N+gM_{\sigma})x\label{eq:split_continuous_sneeze}}
converge to zero  as fast as $\text{exp}({-\lambda t})$ does.\end{lemma}
 
 \vspace{-0.1in}
 
 The proofs of Lemma \ref{lem:split_continuous_average} and   Theorem \ref{thm:split_continuous_average} can be found in Appendix A.
  
 \vspace{-0.1in}

 In the sequel the problem is studied for a certain type of switching neighbor graphs. 
 It turns out that if the stochastic matrices of undirected neighbor graphs are chosen to be doubly stochastic, there exist estimators which can deal with arbitrary switching signals,  and the notion of dwell times ceases to be relevant.
 The estimator to be considered is  again the same as \eqref{eq:split_continuous_observer}, with the exception  that $g$ is chosen differently.

 \vspace{-0.1in}
 
 \begin{theorem} 
 For any fixed positive number $\lambda$,
 there exists a positive number $g^*$ with the following property.
 For any  value of $g\geq g^*$, any time-varying neighbor graph $\mathbb{N}(t)$, if the system defined by \eqref{eq:pre_sys1} and \eqref{eq:pre_sys2} is jointly observable, the neighbor graph $\mathbb{N}(t)$ is undirected and connected, and  the stochastic matrix $S(t)$ of graph $\mathbb{N}(t)$ is doubly stochastic, there are matrices $K_i,\;i\in\mathbf{m}$ such that  each state estimation error $x_i(t)-x(t) $ of the  distributed observer defined by \eqref{eq:split_continuous_observer},
 converges to zero
as  $t\rightarrow \infty$ as fast as $\text{exp}({-\lambda t})$ converges to zero. \label{thm:arb}\end{theorem}

 \vspace{-0.1in}
 
 The proof of Theorem \ref{thm:arb} can be found in Appendix~A.

Given that $\tilde A$ has eigenvalues which are a subset of those of $A$, it is seen that the effect of large $g$ is to force the instantaneous value of the eigenvalues of $A_V(t)$ well to the left of those of $\tilde A$, and indeed the same for the Lyapunov exponent.  This is a spectral separation idea -- consensus dynamics within the estimator are faster than those of the original system.

 \vspace{-0.1in}

 \subsection{Distributed Estimator with Adaptive Gains}\label{sec:adaptive_gain}

 \vspace{-0.1in}
 
 Obviously it may be disadvantageous to share a gain across the whole network, and here we
 %\cite{Kim20} has examined how one might find different gains for each estimator in a distributed way. Here we 
%  EITHER: EXPLAIN [22} or {PRESENT A WAY PERMITTING THIS TO BE DONE USING AN ADAPTIVE CONTROL APPROACH. }
%  It has been noticed that there are literature  \cite{Kim20} on how to find suitable gains in a distributed way  for the estimators.
% In this section, we 
aim to design a simple adaptive distributed estimator to get gains for each estimators in a distributed way. 
The estimator for each agent $i$ still  has the form \eqref{eq:split_continuous_observer} while each agent $i$ has its own gain $g_i$ which  is obtained by
\vspace{-0.1in}
{\small \begin{equation}
\dot g_i=\Big|V_i'\sum_{j\in\mathcal N_i} \frac{1}{m_i}
(x_j-x_i)\Big|_2^2,\;\;\;i\in\mathbf m
\label{eq:adaptive}
\end{equation}}
where $|\cdot|_2$ denotes the two norm of a matrix and $g_i(0)$ is nonnegative but otherwise arbitrary. 
 Key questions arising are whether the $g_i$ are bounded, and whether the matrices $K_i$ can be chosen in the same way as previously. We have in fact with $K_i$  chosen as in the proof of Proposition \ref{prop:error}:

 \vspace{-0.1in}

 \begin{theorem}\label{thm:split_continuous_adaptive}
For  any neighbor graph $\mathbb{N}\in\{\mathbb{G}_{\sigma}:\sigma\in\mathcal{S}_{\text{avg}}\}$,  if the system defined by \eqref{eq:pre_sys1} and \eqref{eq:pre_sys2} is jointly observable, and the gain is defined by \eqref{eq:adaptive}, there are matrices $K_i,\; i\in \mathbf m$ such that all the $g_i$ are bounded, and each state estimation error
 $e_i = x_i-x$   of the distributed estimator defined by $\eqref{eq:split_continuous_observer}$  asymptotically converges to zero as $t\rightarrow \infty$.\end{theorem}
 
 \vspace{-0.1in}
 
 The proof of Theorem \ref{thm:split_continuous_adaptive} can be found in Appendix A.

  As with any adaptive control algorithm, we must recognize that there are fundamental challenges that can arise in practice and have the potential to undermine the approach \cite{Anderson1998}: these include 
the need to work with models of plants that may be very accurate but are virtually never exact; the inability to know, given an unknown plant, whether a desired control objective is practical or impractical, and the possibility of transient instability, or extremely large signals occurring before convergence. 
Thus, for this paper, our preference is to  stick with a given $g$ instead of using an adaptive algorithm.

 \vspace{-0.1in}

 \subsection{Resilience }\label{sec:resilience}
  
 \vspace{-0.1in}
 
 The concept of a passively resilient algorithm is proposed in \cite{Automatica22}.
 By a passively resilient algorithm for a distributed process is meant an algorithm which, by exploiting built-in
network and data redundancies, is able to continue to
function correctly in the face of abrupt changes in the
number of vertices and arcs in the inter-agent communication graph upon which the algorithm depends.
All the proposed continuous-time distributed estimators, except for the adaptive one, are inherently resilient to these abrupt changes provided the network is redundantly strongly connected and redundantly jointly observable, with a careful gain picking before the algorithm starts. Details can be found in Section 5 of \cite{Automatica22}. The same resilience property is also possessed by the discrete-time distributed estimators to be developed in the next section.

%From the preceding subsections we know that exponential state estimation can be achieved even if the neighbor graph is time-varying.  This implies that the proposed algorithm can deal with the time-varying topology due to communication failures when the network redundancy condition is satisfied. Details can be found in Section 5 of \cite{Automatica22}. Thus the algorithm proposed  is inherently resilient to communication failures.
\vspace{-0.1in}
 \section{Discrete-Time Distributed Estimator}\label{sec:split_discrete}
 \vspace{-0.1in}
 In this section, a discrete time version of the distributed estimator problem is studied. 
 The estimator which solves this problem in discrete time is described.   
Of central concern is the achieving of the split spectrum property, which underpins a stability guarantee for the estimator. High gain is not the answer: in continuous time, this produced a split spectrum (with some very fast modes). 
 The discrete-time analogy rests on having part of the spectrum in the estimator achieved by gains $K_i$ as in continuous-time and partly determined by the dynamics creating consensus among the components of local estimated states corresponding to unobserved components of the underlying system state, with the associated time scale significantly faster than the single time scale associated with the underlying system dynamics. The second time scale is made possible through the introduction of a faster sampling rate.

% Several techniques are outlined for picking the number of switches required between ``event times''  in order to achieve  a prescribed convergence rate.
 
% \subsection{Problem Formulation}
\vspace{-0.1in}
 
  We are interested in the same time-varying network as used for the continuous-time linear system and which is characterized by the neighbor graph  $\mathbb{N}(t)$. 
  Each agent $i$ can sense a discrete-time  signal $y_i(\tau)\in\R^{s_i}$ at {\em event times} $\tau T$, $\tau = 0, 1,2,\ldots $  where $T$ is a positive constant; for  $i\in\mathbf{m}$ and $\tau=0,1,2,\ldots $
\eq{y_i(\tau)   =  C_ix(\tau) ,\;\;\;\;\;x(\tau +1) =Ax(\tau )
\label{eq:split_syss_d}} and $x\in\R^n$. 
We assume throughout that $\mathbb{N}(t)$ is strongly connected, and the system defined by
\rep{eq:split_syss_d} is {\em jointly observable}.

Each agent $i$ is to  estimate $x$ using a dynamical system whose
output  $x_i(\tau)\in\R^n$
 is to be an asymptotically correct estimate of $x(\tau)$ in the sense that the estimation error
$x_i(\tau)-x(\tau)$ converges to zero  as $\tau\rightarrow \infty$
as fast as $\lambda^{\tau} $ does,
 where $\lambda$ is an
arbitrarily chosen but fixed
 positive number\footnote{For the type of observer to be developed, finite-time convergence is not possible.} less than $1$. To accomplish this it is assumed that
 the information
agent $i$ can receive from neighbor $j$ at event time  $\tau T$ is
$x_j(\tau)$. It is further
 assumed that agent $i$
can also receive certain additional information from its neighbors
at a finite number of
 times  between each successive pair of  event times; what this  information is  will be specified below.

\vspace{-0.1in}
\subsection{The Estimator}\label{chap:split-discrete-obs} 
\vspace{-0.1in}
In this section it will be assumed that each agent's neighbors do not
 change between event times. In other words,  for $i\in\mathbf{m}$,  $$\mathcal{N}_i(t) =
  \mathcal{N}_i(\tau T),\;\;\;\; t\in [\tau T,(\tau+1)T),\;\;\;\;\tau = 0,1,2,\ldots $$
With this assumption, the estimator to be considered
 consists of $m$ private estimators, one for each agent.
  The estimator  for agent $i$ is  of the form
\eq{ x_i(\tau+1) =
 (A+K_iC_i)\bar{x}_i(\tau) -K_iy_i(\tau)\label{eq:split-discrete-est1}}
where $\bar{x}_i(\tau)$ is an ``averaged state'' computed
recursively  over $q$ steps  during the real time interval $[\tau T,\; (\tau+1)T)$
using the update  equations
\begin{eqnarray}
z_i(0,\tau) &=&x_i(\tau)\label{eq:split-discrete-p1}\\
z_i(k,\tau) &=& (I-P_i)z_i(k-1,\tau)+ \nonumber\\ &\ & \frac{1}{m_i(\tau)}P_i\!\!\!\!\sum_{j\in\mathcal{N}_i(\tau T)}\!\!\!\!z_j(k-1,\tau),\;\;k\in \mathbf{q}\\
\bar{x}_i(t) &= &z_i(q,\tau)\label{eq:split-discrete-p3}\end{eqnarray} Here
$m_i(\tau)$ is the number of labels in $\mathcal{N}_i(\tau T)$, $q$ is a
suitably defined positive integer,  further detail being given below,
 $\mathbf{q}=\{1,2,\ldots,q\}$,  and   $P_i$ is the orthogonal projection on the unobservable space of $(C_i,A)$.
Each matrix  $K_i$ is defined as explained   in the next paragraph. Meanwhile, we note that the estimators incorporate two time scales. An agent's local estimator obtains data from its neighbors at a rate $q$ times that at which it obtains measurement data from the underlying system.  

 As  described in \S \ref{sec:split_continuous_estimator},   for  fixed $i\in\mathbf{m}$,  write $Q_i$ for any full  rank matrix whose kernel
    is the unobservable space of $(C_i,A)$, and let $\bar{C}_i$ and $\bar{A}_i$ be the unique solutions to
    $\bar{C}_iQ_i = C_i$ and $Q_iA=\bar{A}_iQ_i$  respectively.
    Let $\lambda$ be a positive value bounded by $1$.
    Then  the matrix pair $(\bar{C}_i,\bar{A}_i)$ is
     observable.  Thus by using a standard spectrum assignment algorithm,
     a matrix $\bar{K}_i$ can be chosen to ensure that  the convergence
     of $(\bar{A}_i + \bar{K}_i\bar{C}_i)^{\tau}$ to zero as $\tau\rightarrow \infty$
     is as fast as the convergence to zero
      of  $\lambda ^{\tau}$.
      Having chosen such $\bar{K}_i$, $K_i$
      is then defined to be $K_i = Q_i^{-1}\bar{K}_i$ where $Q_i^{-1}$ is a right inverse for $Q_i$.
% The definition implies that $Q_i(A+K_iC_i) =
% (\bar{A}_i+\bar{K}_i\bar{C}_i)Q_i$ and that
% $(A+K_iC_i)\mathcal{V}_i\subset \mathcal{V}_i$. And there is a unique matrix  $A_i$ which satisfies $(A+K_iC_i)V_i
% = V_iA_i$ where $V_i$ is a basis matrix for $\mathcal{V}_i$.
% % Note in addition that
%         %$\sigma(A+K_iC_i) = \sigma(\bar{A}_i +\bar{K}_i\bar{C}_i)\cup\sigma{A_i}$.
        To explain what needs to
         be considered in choosing
        $q$, which is a rough analog of the gain $g$ of the continuous-time solution, it is necessary to describe  the structure of the  ``error model'' of  the overall estimator. This will be done next.

\vspace{-0.1in}
\subsubsection{The Error Dynamics}
\vspace{-0.1in}
For $i\in\mathbf{m}$, write $e_i(\tau)$  for the {\em state
estimation error} $e_i(\tau) = x_i(\tau)-x(\tau )$. In view of
\rep{eq:split-discrete-est1},
$$e_i(\tau+1) = (A+K_iC_i)\bar{e}_i(\tau)$$
where $\bar{e}_i(\tau) =\bar{x}_i(\tau)-x(\tau)$.  Moreover if  %$\epsilon_i(0,t) =e_i(t)$ and
 $\epsilon_i(k,\tau) \dfb z_i(k,\tau)-x(\tau),\;k\in\{0,1,\ldots, q\}$
then for $k\in \mathbf{q}$,
\begin{eqnarray*}
\epsilon_i(0,\tau) & =& e_i(\tau)\\
\epsilon_i(k,\tau) &=& (I-P_i)\epsilon_i(k-1,\tau)\!+ \!\frac{1}{m_i(\tau)}P_i\!\!\!\!\!\sum_{j\in\mathcal{N}_i(\tau T)}\!\!\!\!\!\epsilon_j(k-1,\tau\!)\\
\bar{e}_i(\tau) &= &\epsilon_i(q,\tau)\end{eqnarray*} because of
\rep{eq:split-discrete-p1} -- \rep{eq:split-discrete-p3}. It is possible to combine   these $m$
subsystems  into a single system. Paralleling \S \ref{sec:error_continuous} let
  $e = $ col $\{e_1,e_2,\ldots, e_m\}$, define
$\bar{A} = $ block diag $\{A+K_1C_1
,A+K_2C_2,\ldots,A+K_mC_m\}$,
 $P = $ block diag $\{P_1 ,P_2,\ldots, P_m\}$ and write $S(\tau)$ for the stochastic matrix
$S(\tau) = D_{\mathbb{N}(\tau T)}^{-1}\mathcal A'_{\mathbb{N}(\tau T)}$ where
$\mathcal A_{\mathbb{N}(\tau T)}$ is
 the adjacency matrix of $\mathbb{N}(\tau T)$ and  $ D_{\mathbb{N}(\tau T)}$
 is the diagonal matrix whose $i$th diagonal entry is the in-degree
   of $\mathbb{N}(\tau T)$'s $i$th vertex. %Note that $\mathbb{N}(\tau T )$ is the graph
%  of $S'(\tau)$ and that the diagonal entries of $S'(\tau)$
%  are all positive because each agent is a neighbor of itself.
Let $\bar{e}(\tau) = $ col
$\{\bar{e}_1(\tau),\bar{e}_2(\tau),\ldots, \bar{e}_m(\tau)\}$ and
$\epsilon(k,\tau) =
\text{col}\{\epsilon_1(k,\tau),\epsilon_2(k,\tau),\ldots
,\epsilon_m(k,\tau)\}$. Then
$$e(\tau +1) = \bar{A}\bar{e}(\tau)$$
 and
\begin{eqnarray*}
\epsilon(0,\tau) &=&e(\tau)\\
\epsilon(k,\tau) &=& (I_{mn}-P(I_{mn}-\bar{S}(\tau)))\epsilon(k-1,\tau),\;k\in\mathbf{q}\\
\bar{e}(\tau) &= &\epsilon(q,\tau)\end{eqnarray*} where
$\bar{S}(\tau ) = S(\tau)\otimes I_n$. Clearly
$\bar{e}(\tau) = (I_{mn}-P(I_{mn}-\bar{S}(\tau)))^qe(\tau)   $,
so \eq{e(\tau+1) = \bar{A}(I_{mn}-P(I_{mn}-\bar{S}(\tau)))^qe(\tau)
\label{eq:split-discrete-error}}

Our immediate aim is now to explain why  for $q$ sufficiently large, the
time-varying matrix
 $\bar{A}(I_{mn}-P(I_{mn}-\bar{S}(\tau)))^q$ appearing in \rep{eq:split-discrete-error} is a discrete-time stability matrix
for which the product \eq{\Phi(\tau) = \prod_{s =
1}^{\tau}\bar{A}(I_{mn}-P(I_{mn}-\bar{S}(s)))^q\label{eq:split-discrete-phi}}
converges to zero as $\tau\rightarrow\infty $ as fast as $\lambda
^{\tau}$ does. 

To begin with, we explore the propoerty of matrix $\bar A (I_{mn}-P(I_{mn}-\bar S(\tau)))^q$.
\vspace{-0.1in}

\begin{proposition}\label{prop:errordis}
The spectrum of the error system matrix $\bar A (I_{mn}-P(I_{mn}-\bar S(\tau)))^q$ splits into two subsets.  One subset contains  the union of certain subsets of the eigenvalues associated with the $i$-th local estimator, $i\in\mathbf m$, these being  able to  be arbitrarily positioned by choice of the $K_i$ and independent of $q$. The second subset is  independent of the choice of the $K_i$ and depends, though not to the extent of being able to be arbitrarily positioned, on~$q$.
\end{proposition}
\vspace{-0.1in}
{ \bf Proof of Proposition \ref{prop:errordis}:}
As described in \S \ref{sec:split_continuous_estimator}, note that  the
subspace $\mathcal{V} = \mathcal{V}_1\oplus\mathcal{V}_2\oplus \cdots \oplus
\mathcal{V}_m$
 is
$\bar{A}$ - invariant  because $(A+K_iC_i)\mathcal{V}_i\subset
\mathcal{V}_i,\;i\in\mathbf{m}$. Next, let $Q = $ block diag  $
\{Q_1, Q_2,\ldots ,Q_m\}$ and $V = $ block diag  $ \{V_1,
V_2,\ldots ,V_m\}$  
 It follows that $P = VV'$, \eqref{eq:split_QA} and \eqref{eq:split_AV}. 
Also as before, $(A+K_iC_i)V_i=V_iA_i$. Moreover
\begin{eqnarray}Q(I_{mn}-P(I_{mn}-\bar{S}(\tau)))^q &=& Q\label{eq:split_discrete_pr1}\\
(I_{mn}-P(I_{mn}-\bar{S}(\tau)))^qV& =& V(
V'\bar{S}(\tau)V)^q\label{eq:split_discrete_pr2}\end{eqnarray} Note that \rep{eq:split_discrete_pr1}
holds  because $QP = 0$.  To understand why \rep{eq:split_discrete_pr2} is true, note
first that $(I_{mn}-P(I_{mn}-\bar{S}(\tau)))V =  V(I_{\bar{n}} -
V'(I_{mn} - \bar{S}(\tau))V)$ because $P =VV'$; here $\bar{n} =
\dim(\mathcal{V})$. But $I_{\bar{n}} - V'(I_{mn} - \bar{S}(\tau))V =
V'\bar{S}(\tau)V$ because
  $V'V = I_{\bar{n}}$. Thus \rep{eq:split_discrete_pr2} holds for $q=1$; it follows by induction that \rep{eq:split_discrete_pr2}
  holds for any positive integer $q$.

Using \eqref{eq:split_QA}, \eqref{eq:split_AV},\eqref{eq:split_discrete_pr1}, and \eqref{eq:split_discrete_pr2}, one obtains the equations
\begin{eqnarray}Q\bar{A}(I_{mn}-P(I_{mn}-\bar{S}(\tau)))^q &=& \bar{A}_VQ\label{eq:split_discrete_xpr1}\\
\bar{A}(I_{mn}-P(I_{mn}-\bar{S}(\tau)))^qV& =& VA_V(\tau)
\label{eq:split_discrete_xpr2}\end{eqnarray} where \eq{A_V(\tau) =
\tilde{A}(V'\bar{S}(\tau)V)^q\label{eq:split_discrete_ench}} 
with $\tilde A=\text{block diag }\{A_1,A_2,\ldots,A_m\}$.
These equations imply
that{\footnote{The notation $\hat A_V$ and  $A_V$  are different from the two defined  in \S \ref{sec:split_continuous}.}}
{\small\eq{\bar{A}(I_{mn}-P(I_{mn}-\bar{S}(\tau)))^q = T\begin{bmatrix}\bar{A}_V & 0\\
\hat{A}_V(\tau) & A_V(\tau)\end{bmatrix}T^{-1} \label{eq:split_discrete_as}}}
with $T=\begin{bmatrix}
Q^{-1}& V
\end{bmatrix}$ and $\widehat{A}_V(\tau) =
V^{-1}\bar{A}(I_{mn}-P(I_{mn}-\bar{S}(\tau)))^qQ^{-1}$.

According to \eqref{eq:split_discrete_as}, the spectrum of $\bar{A}(I_{mn}-P(I_{mn}-\bar{S}(\tau)))^q$ is equivalent to the union of the spectrum of $\bar A_V$ and $A_V(\tau)$.\hfill $\qed$

% Since the spectrum of each
% $\bar{A}_i+\bar{K}_i\bar{C}_i,\;i\in\mathbf{m}$, is assignable with
% $\bar{K}_i$, and $\widehat{A}_V(\tau)$  is a bounded matrix, to show
% that  for suitably defined $\bar{K}_i$ and  $q$ sufficiently large,
% the matrix $\Phi(\tau )$ defined in \rep{eq:split-discrete-phi}
%  converges to zero as fast as $\lambda^{\tau}$ does, it is sufficient to show that for $q$ sufficiently large,
%  $A_{V}(\tau )$ is a discrete-time stability matrix whose state-transition matrix converges to zero
% as fast as $\lambda^{\tau}$ does. {\color{red} It is of course obvious that the determinant of $\tilde A(V'\bar S(\tau)V)^q$ goes to zero as $q$ increases, but establishing a corresponding property for the transition matrix is more work. The intuition is that with a sufficiently large $q$, for any $\bar S(\tau)$, the spectral radius of $\bar A(V'\bar S(\tau)V)^q$ will be smaller, probably significantly smaller, than that of $\bar A$, and this spectral separation underpins the stability of the time-varying system associated with changing $\bar S(\tau)$.   
\vspace{-0.1in}
 \subsection{Time-varying Neighbor Graph}
\vspace{-0.1in}
The following result can be obtained when the neighbor graph is time-varying but strongly connected.

\begin{theorem}\label{thm:split_discrete}
 For any given  $\lambda$ with $|\lambda|<1$, if the neighbor graph  $\mathbb{N} (\tau)$  is  strongly connected, and the system defined by \eqref{eq:split_syss_d} is jointly observable, there are matrices $K_i$, $i\in \mathbf{m}$ such that for  sufficiently large $q$,
 each estimation error $x_i(\tau)-x(\tau)$ of the distributed estimator defined by \eqref{eq:split-discrete-est1}-\eqref{eq:split-discrete-p3}, converges to zero as $\tau \rightarrow \infty$ as fast as $\lambda^\tau $ converges to zero.
 \end{theorem}

The proof of the theorem   involves studying the transition matrix and making use of the following results, with all proofs being contained in Appendix B.

%\begin{proposition}
%For each fixed value of $\tau$, $V'\bar{S}(\tau)V$ is a discrete-time stability matrix
%and
%\eq{(V'\bar{S}(\tau)V)' R(\tau)(V'\bar{S}(\tau)V) - R(\tau) < 0\label{ly}}\label{mp} \end{proposition}
%where $R(\tau) =  \Pi(\tau} \otimes I_n$.

\begin{lemma} Let $M$ be an $m\times m$ row stochastic matrix  which  has a strongly connected graph.
There exists  a diagonal matrix $\Pi_M$ whose diagonal entries are
positive for which the matrix $L_M= \Pi_M- M'\Pi_M M $ is positive
semi-definite;  moreover $L_M\mathbf{1} = 0$ where
  $\mathbf{1}$ is the $m$-vector of $1$s. If, in addition,
 the diagonal entries of $M$ are all positive, then
 the  kernel of $L_M$ is one-dimensional.
 \label{lem:split_discrete_brian} \end{lemma}

\begin{proposition}
For each fixed value of $\tau$, \eq{(V'\bar{S}(\tau)V)'
R(\tau)(V'\bar{S}(\tau)V) - R(\tau) < 0\label{eq:split_discrete_ly}} where $R(\tau )$
is the positive definite matrix, $R(\tau) =  V'(\Pi_{S(\tau)}
\otimes I_n)V$.
\label{prop:split_discrete_mp} \end{proposition}

% {\color{red} The proofs of Lemma \ref{lem:split_discrete_brian} and   Proposition \ref{prop:split_discrete_mp} can be found in Appendix B.}

% Note that \rep{eq:split_discrete_ly} shows that for each fixed $\tau$, $x'R(\tau)x$  a
% discrete-time Lyapunov function for the equation $w(k+1)=V'\bar
% S(\tau)Vw(k)$.  Thus for fixed $\tau$,
%   $V'\bar{S}(\tau)V$ is a discrete-time stability matrix.

\begin{remark}

%      I would like to see a
% clear discussion of which steps in the design process require a
% centralized design; also, the computational complexity of this process
% (even if done offline) needs some discussion. 
It should be noticed that the computation of certain gains ($g$ for the continuous-time case, and $q$ for the discrete-time case) requires a centralized design. Besides this, all other design steps are distributed.
Even though  the computation of certain gains 
requires going over all possible strongly connected directed graphs on
$m$ vertices, which is a computationally intensive step,  a clear distinction needs to be drawn between the computations required for designing the algorithm, and those required to run it. 
In design, we can afford to do more computations.
% From the design perspective, the computations would not affect the performance of the proposed distributed ``online'' algorithm.
\end{remark}

% \subsection{Choosing q}
% \label{chap:split-discrete-switch}
% In what follows it will be assumed that each $\bar{K}_i$ has been
% selected so that  the the matrix
%  $\bar{A}_V$ defined by \rep{eq:split_discrete_sunday}, is such that $\bar{A}_V^{\tau}$ converges
%   to zero  as $\tau\rightarrow \infty$ as fast as $\lambda^{\tau}$ does.
%  This can be done using standard
% spectrum assignment techniques to  make the spectral radius of
% $\bar{A}_V$ at least as small as $\lambda $.
%  In view of   \rep{eq:split_discrete_as}, it is clear that to assign the convergence rate of the state transition matrix of
% $\bar{A}(I_{mn}-P(I_{mn}-\bar{S}(\tau)))^q$ it is necessary and
% sufficient to control the convergence rate
%  of the state transition matrix of $A_{V}(\tau)$. This, {\color{cyan} as we will now show,} can be accomplished by choosing $q$ sufficiently large.
% {\color{cyan} We will actually detail} two different ways to do this, each utilizing a different
% matrix norm. Both approaches
%  will be explained next
% using the abbreviated notation  $B(\tau) = V'\bar{S}(\tau)V$; note
% that with this simplification, $A_V(\tau) = \tilde{A}B^q(\tau)$
% because of \rep{eq:split_discrete_ench}.

%  \subsection{Comparison}
 
 \vspace{-0.1in}

\section{Simulations}\label{sec:simu}
 
 \vspace{-0.1in}
 
This section provides simulations to illustrate the state estimation performance for both continuous and discrete time systems. 
%Simulation results for time varying networks are also provided. 
The neighbor graph  in some simulations will switch back and forth between Fig.~\ref{graph} (a) and Fig.~\ref{graph} (b).  On occasion, it can serve to model  a connection failure happening between agent $1$ and agent $3$ randomly.
 
 \vspace{-0.1in}

\subsection{Continuous dynamics}\label{sec:simucontinuius}
 
 \vspace{-0.1in}
 
Consider the three channel, four-dimensional, continuous-time system
described by the equations  $\dot x=Ax,\; y_i=C_ix,\; i\in \{1,\;2,\;3\}$, where
 
 \vspace{-0.1in}
{\small $$A=\begin{bmatrix} 0& 1& 0 & 0\\ -1&0&0&0\\ 0 &0 & 0 & 1\\
0 & 0 & -2 & 0\end{bmatrix} $$} and $C_i$ is the $i$th unit row
vector in $\R^{1\times 4}$. Note that $A$ is a  matrix with
eigenvalues at $\pm 1 j$, and $\pm1.4142 j$.
 While the system is jointly observable, no single pair $(C_i,A)$  is observable. 
   The   local observer convergence rate is designed to be at least with rate $\lambda=1$. The first step is to design $K_i$ as stated in \S \ref{sec:split_continuous_estimator}. { This is to control  the spectrum of the matrix $\bar A_V$ as defined in \eqref{eq:split_continuous_split}.}
   
    \noindent For agent 1:  \begin{eqnarray*} &&
 {A}_1=\begin{bmatrix}0 & -1\\ 1 & 0\end{bmatrix},\;\;\; {Q}_1=\begin{bmatrix}0&1& 0 & 0\\ 1& 0 & 0 &  0\end{bmatrix},\;\;V_1 = \begin{bmatrix}
0 & 0 & 1 & 0\\ 0 & 0 & 0 & 1
\end{bmatrix}',\\ &&K_1 =
\begin{bmatrix}-5 & -5& 0& 0
\end{bmatrix}'\end{eqnarray*} \noindent For
agent 2:
  \begin{eqnarray*} &&
 {A}_2=\begin{bmatrix}0 & -1\\ 1 & 0\end{bmatrix},\;\; {Q}_2=\begin{bmatrix}-1&0 & 0 & 0\\ 0 & 1 & 0 &  0\end{bmatrix},\;\;V_2 = \begin{bmatrix}
0 & 0 & 1 & 0\\ 0 & 0 & 0 & 1
\end{bmatrix}',\\ && K_2 =
\begin{bmatrix}5& -5& 0 & 0
\end{bmatrix}'\end{eqnarray*} \noindent
For agent 3:     \begin{eqnarray*} &&
 {A}_3=\begin{bmatrix}0 & -2\\ 1 & 0\end{bmatrix},\;\; {Q}_3=\begin{bmatrix}0&0 &0 & 1\\ 0 &0 & 1 &  0\end{bmatrix},\;\;V_3 = \begin{bmatrix}
1 & 0 & 0 & 0\\ 0 & 1 & 0 & 0
\end{bmatrix}',\\ && K_3 =
\begin{bmatrix}0 & 0& -5& -4
\end{bmatrix}'\end{eqnarray*}

 \begin{figure}[ht]
\centerline{\includegraphics [ height =1.4 in]{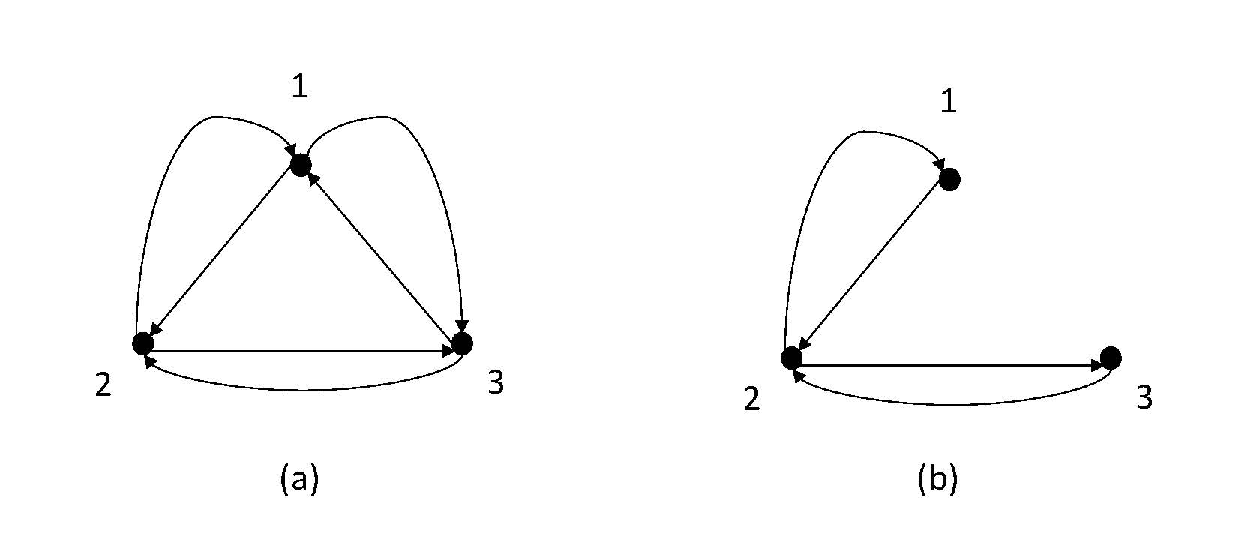}}
\caption{The neighbor graph}
   \label{graph}
\end{figure}

\begin{figure}
     \centering
     \begin{subfigure}[b]{0.2\textwidth}
         \centering
         \includegraphics[width=\textwidth]{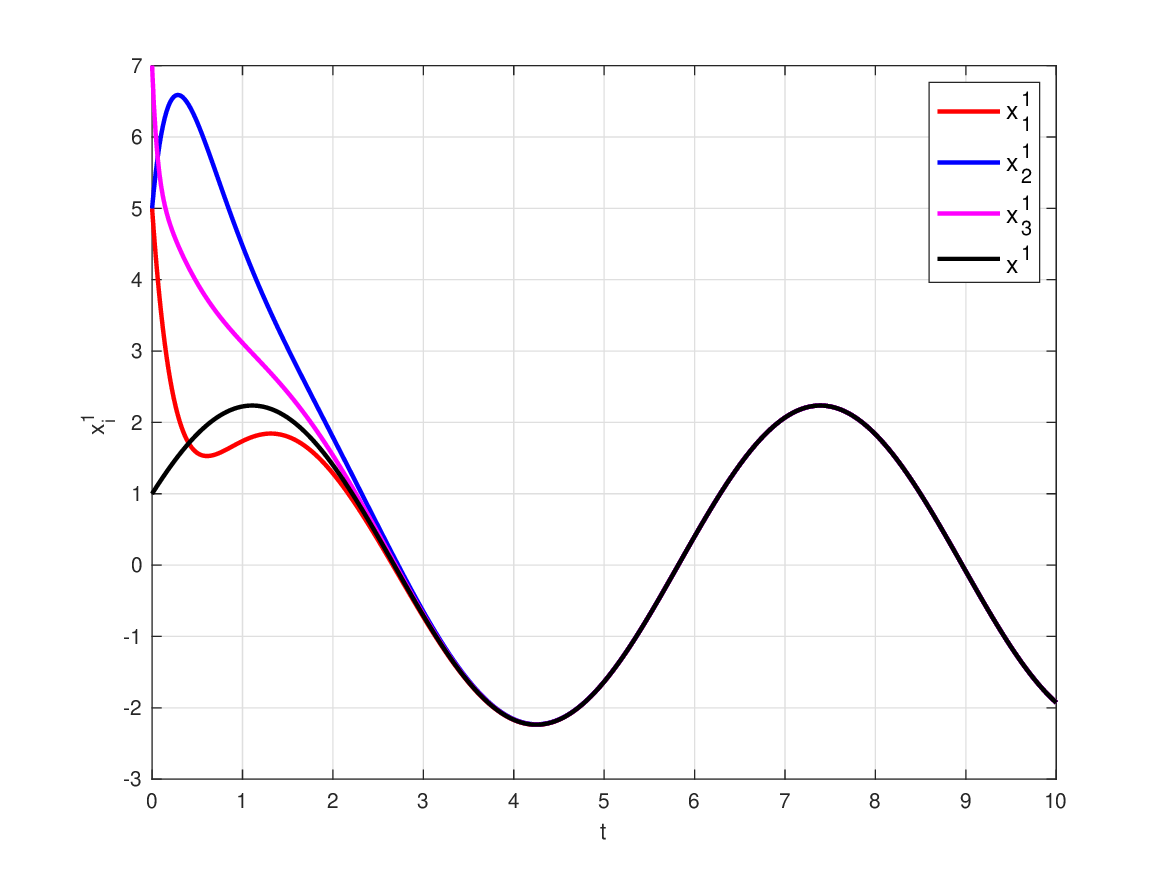}
         \caption{Trajectory of the performance}
         \label{fig:fix1}
     \end{subfigure}
     \hfill
     \begin{subfigure}[b]{0.2\textwidth}
         \centering
         \includegraphics[width=\textwidth]{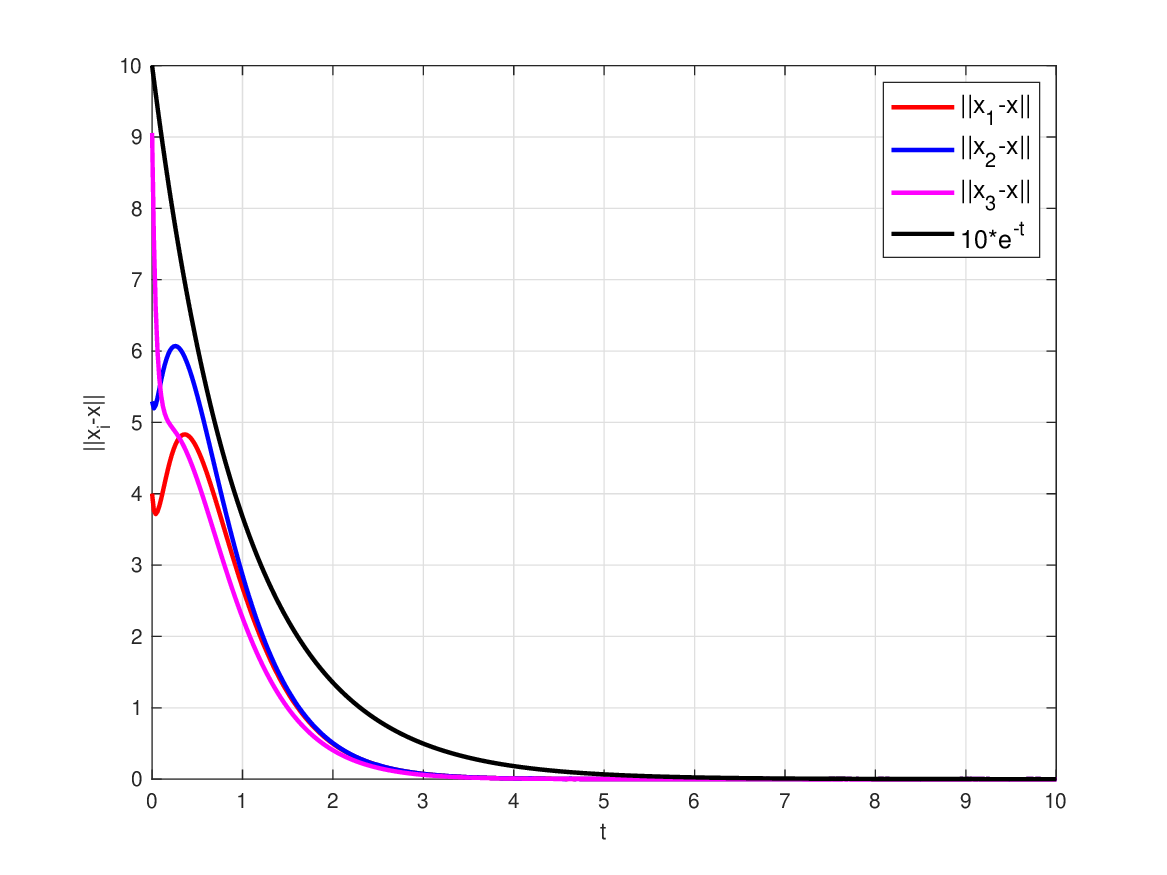}
         \caption{The trajectory of the norm of the estimation error }
         \label{fig:fix2}
     \end{subfigure}
        \caption{The trajectory under time-varying neighbor graphs}
\end{figure}

Two cases are considered.
First, suppose the neighbor graph
$\mathbb{N}(t)$ is fixed as shown in   Figure  \ref{graph}(a). 
 The system considered includes input white noise  with zero mean, that is $\dot x=Ax +v$ where  $E[v(t)]=0, E[v(t)v'(s)]=0.5^2\delta(t-s)$.
    With $g=10$ obtained using \eqref{eq:g},  the real part of the  right most eigenvalue of $A_V$ is less than $-1$. With randomly chosen initial state values, traces of this simulation are shown in Fig.~\ref{fig:fix1} where $x_i^1$ and $x^1$ denote the first components of $x_i$ and $x$ respectively. 
 Moreover, the
norm of the estimation error  is plotted in Fig. \ref{fig:fix2} from
which we can see that it is exponentially convergent with the
  approximate rate $\lambda=1$.

%  \begin{figure}[ht]
% \centerline{\includegraphics [ height =1.6 in]{fix1.jpg}}
% \caption{Trajectory of the performance for systems with noise}
%    \label{fig:fix1}
% \end{figure}

%  \begin{figure}[ht]
% \centerline{\includegraphics [ height =1.65 in]{fixnorm.jpg}}
% \caption{The trajectory of the norm of the  estimation error for systems with noise}
%    \label{fig:fix2}
% \end{figure}

\begin{figure}[ht]
\centerline{\includegraphics [ height =1.3in]{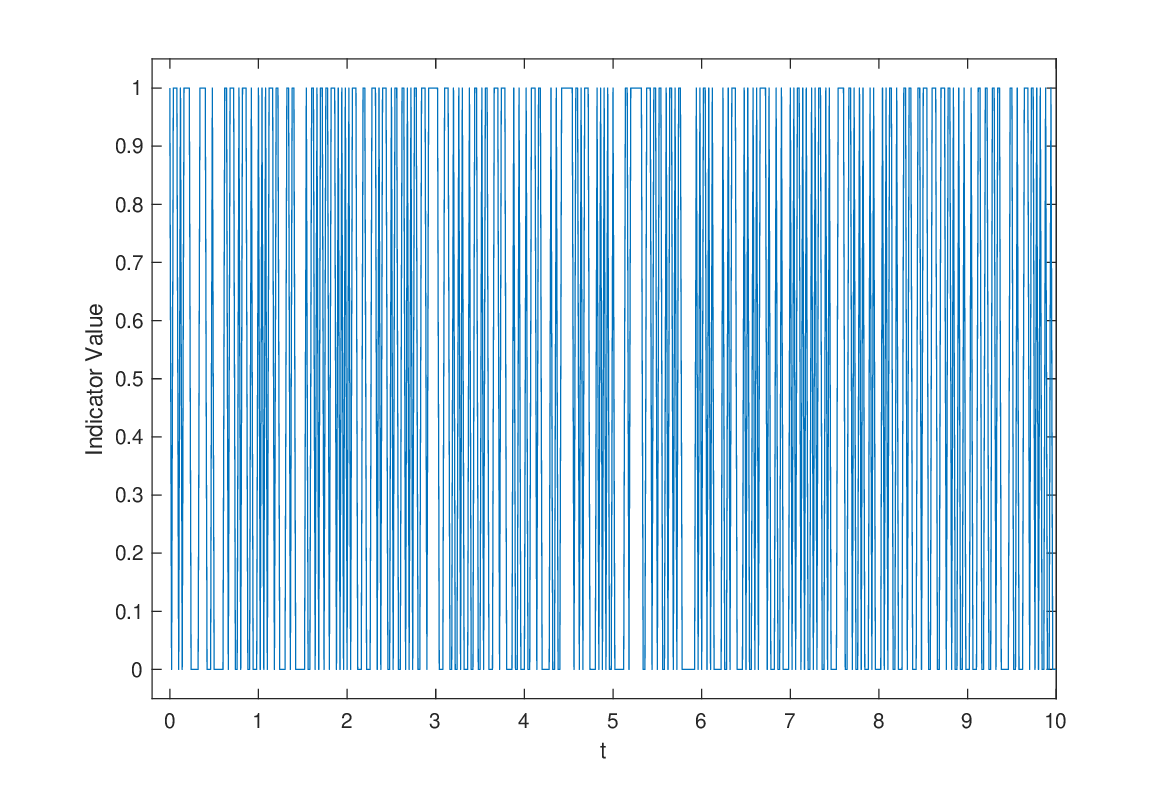}}
\caption{The indicator function}
   \label{fig:indicator}
\end{figure}

\begin{figure}
     \centering
     \begin{subfigure}[b]{0.2\textwidth}
         \centering
         \includegraphics[width=\textwidth]{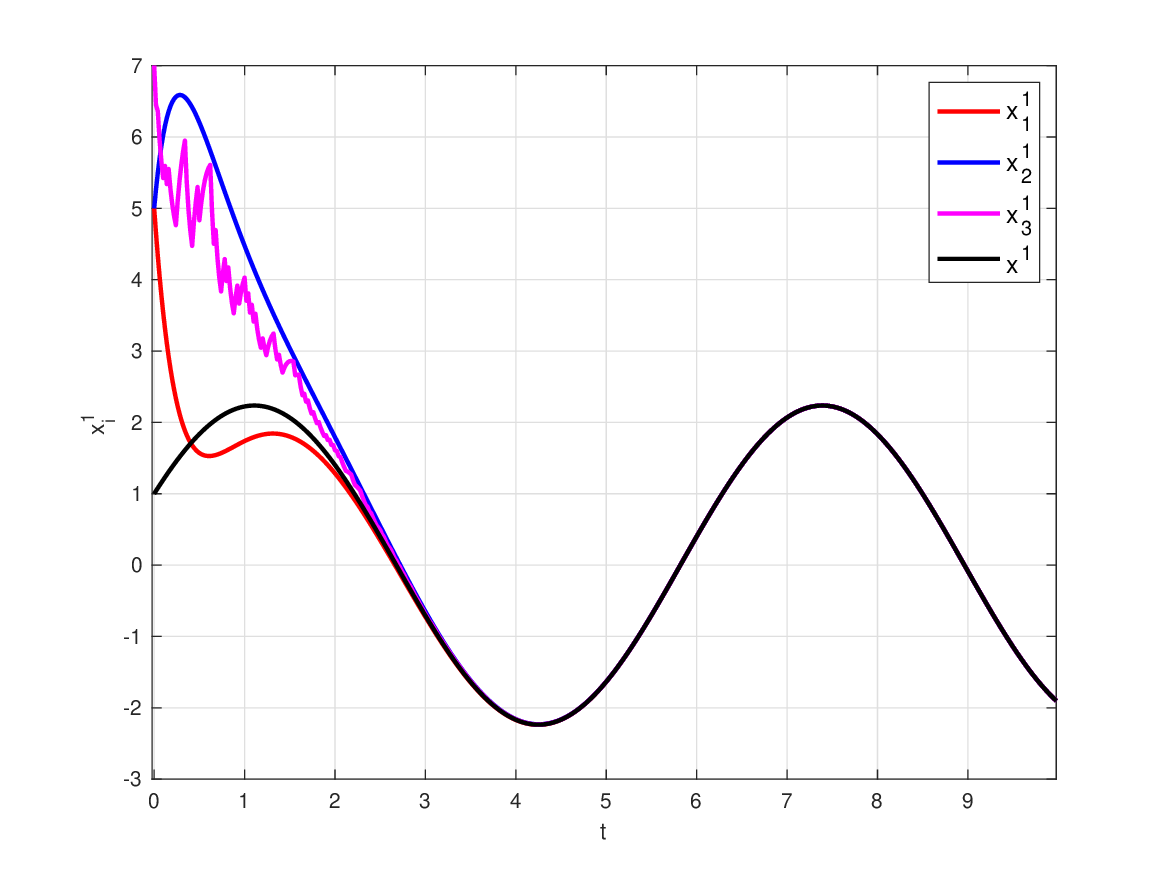}
         \caption{The state value}
         \label{fig:varying}
     \end{subfigure}
     \hfill
     \begin{subfigure}[b]{0.2\textwidth}
         \centering
         \includegraphics[width=\textwidth]{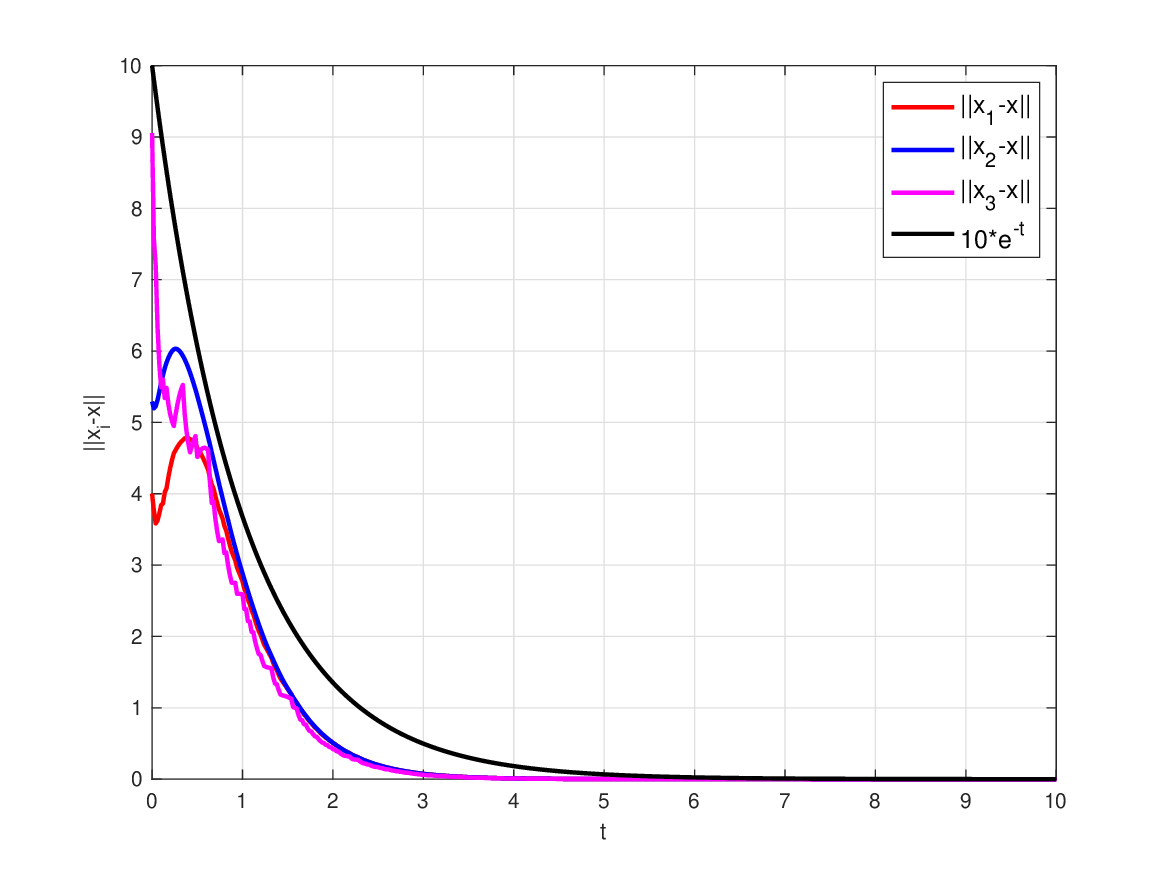}
         \caption{The estimation error }
         \label{fig:varying2}
     \end{subfigure}
        \caption{The trajectory under time-varying neighbor graphs}
\end{figure}

Second, suppose  the neighbor   graph  $\mathbb{N}(t)$ is time-varying and switching back and forth between    Figure  \ref{graph}(a) and Figure \ref{graph}(b) according to the indicator function in Fig.~\ref{fig:indicator}.  That is, when the function value is $1$, the neighbor graph is Figure  \ref{graph}(a), and when the function value is $0$, the neighbor graph is Figure  \ref{graph}(b). It is arranged that the average dwell time is $\tau_D=0.0369$ for this simulation.  
With zero measurement noise the corresponding solution trajectories for  $x_i^3$ and $x^1$  are shown in Fig.~\ref{fig:varying} and the norm of the estimation error is shown in Fig.~\ref{fig:varying2}.

%  \begin{figure}[ht]
% \centerline{\includegraphics [ height =1.75 in]{vayring1.eps}}
% \caption{Trajectory of the performance under time-varying neighbor graphs}
%    \label{fig:varying}
% \end{figure}
%  \begin{figure}[ht]
% \centerline{\includegraphics [ height =1.75in]{varyingerror.eps}}
% \caption{The trajectory of the norm of the estimation error  under time-varying neighbor graphs}
%    \label{fig:varying2}
% \end{figure}

 \vspace{-0.1in}

\subsection{Discrete dynamics}
 
 \vspace{-0.1in}
 
The following simulations are intended to illustrate  how to pick the integer parameter $q$ of the observer.
%the observer performs when applied to an unstable system.
Consider the three channel, four-dimensional, discrete-time system
described by the equations  $x(\tau+1)=Ax(\tau),\; y_i=C_ix,\; i\in \{1,\;2,\;3\}$, where
$A$ and $C_i$ are the same as described in \S \ref{sec:simucontinuius}.
   The  observer convergence rate is designed to be~$\lambda=0.5$. The first step is also to design $K_i$ as stated in \S \ref{chap:split-discrete-obs}.
  For each agent $i$, matrices $A_i$, $Q_i$, and $V_i$ remain the same as stated in \S \ref{sec:simucontinuius}.
  
  \noindent 
  For agent 1:   $K_1 =
\begin{bmatrix}0.5 & 0.94& 0& 0
\end{bmatrix}'$

\noindent For
agent 2:
$K_2 =
\begin{bmatrix}-0.94& 0.5& 0 & 0
\end{bmatrix}'$

\noindent 
For agent 3:  $K_3 =
\begin{bmatrix}0 & 0& 0.5&1.94
\end{bmatrix}'$

\begin{figure}
     \centering
     \begin{subfigure}[b]{0.2\textwidth}
         \centering
         \includegraphics[width=\textwidth]{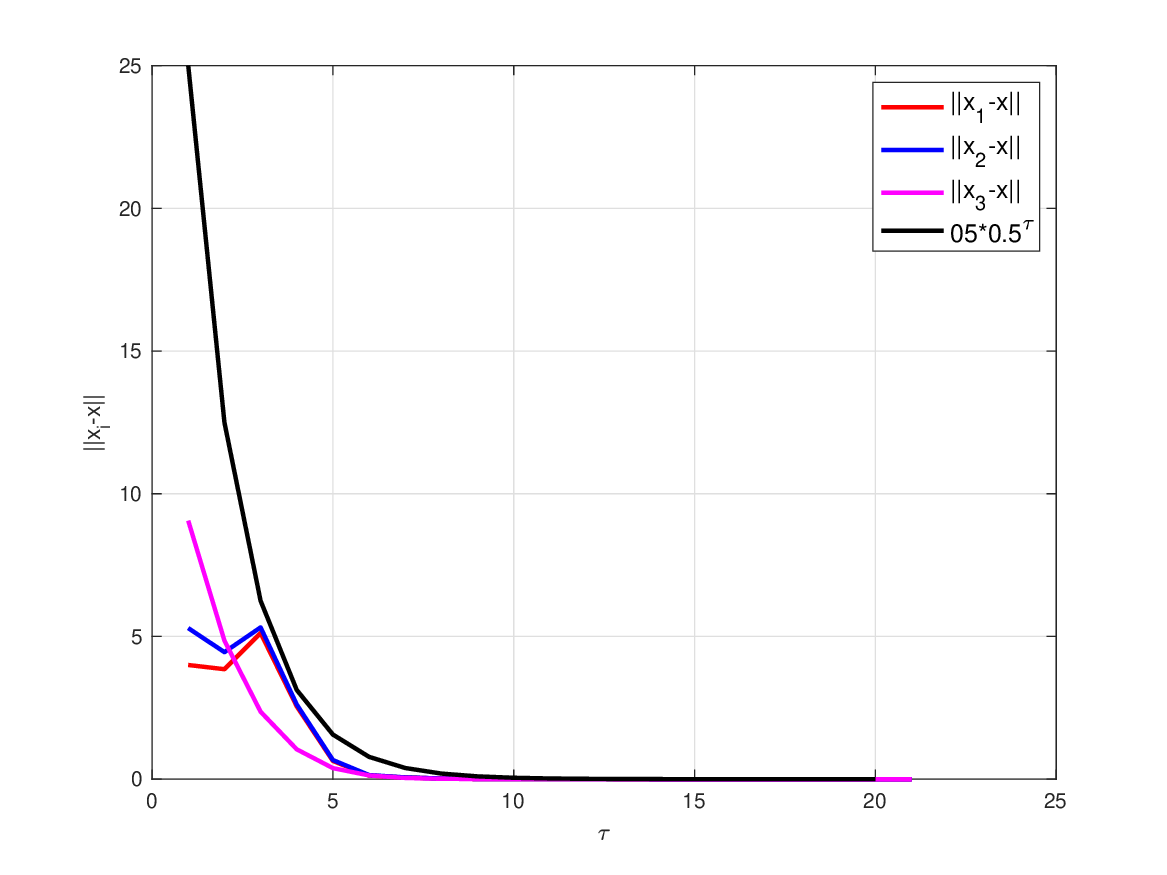}
         \caption{Without noise}
         \label{fig:di-fix1}
     \end{subfigure}
     \hfill
     \begin{subfigure}[b]{0.2\textwidth}
         \centering
         \includegraphics[width=\textwidth]{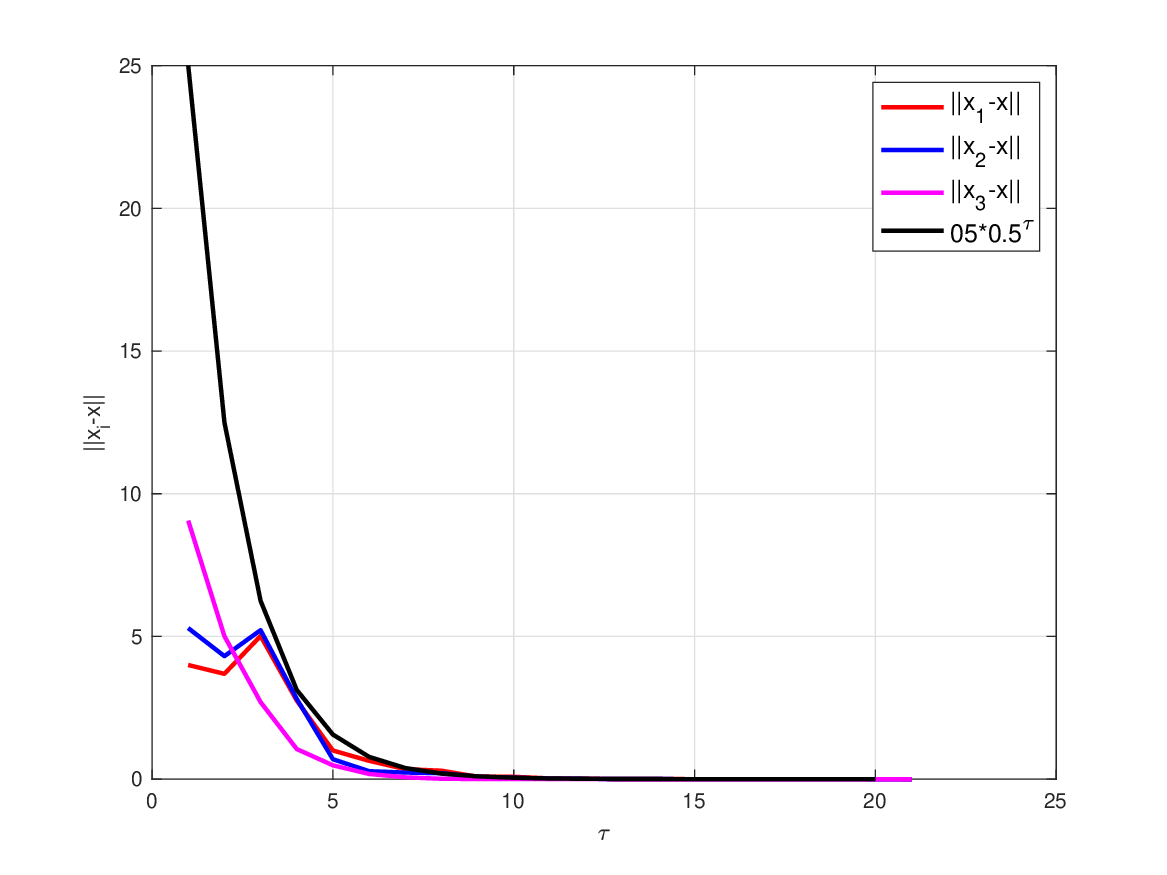}
         \caption{With noise}
         \label{fig:dvaryingerror}
     \end{subfigure}
        \caption{The trajectory of the norm of the estimation error for systems}
\end{figure}

Consider the case when the neighbor graph $\mathbb{N}$ is constant  as in Fig.~\ref{graph} (a).
With randomly chosen initial state values and $q=6$ obtained using \eqref{rajit2},  which leads to a spectral radius of $\tilde A(V'\bar SV)^6$ less than $0.5$,
the norm of the estimation error is plotted in Fig.~\ref{fig:di-fix1} from which we can see that it is exponentially convergent with the rate $\lambda=0.5$. The error traces are bounded by the curve $50\times 0.5^\tau$.
% traces of the simulation is shown in Fig.~\ref{fig:di-fix1} where $x_i^1$ and $x^1$ represent the first components of $x_i$ and $x$ respectively.  

%  \begin{figure}[ht]
% \centerline{\includegraphics [ height =1.7in]{discretefixerror.jpg}}
% \caption{The trajectory of the norm of the estimation error}
%    \label{fig:di-fix1}
% \end{figure}

\begin{figure}[ht]
\centerline{\includegraphics [ height =1.3in]{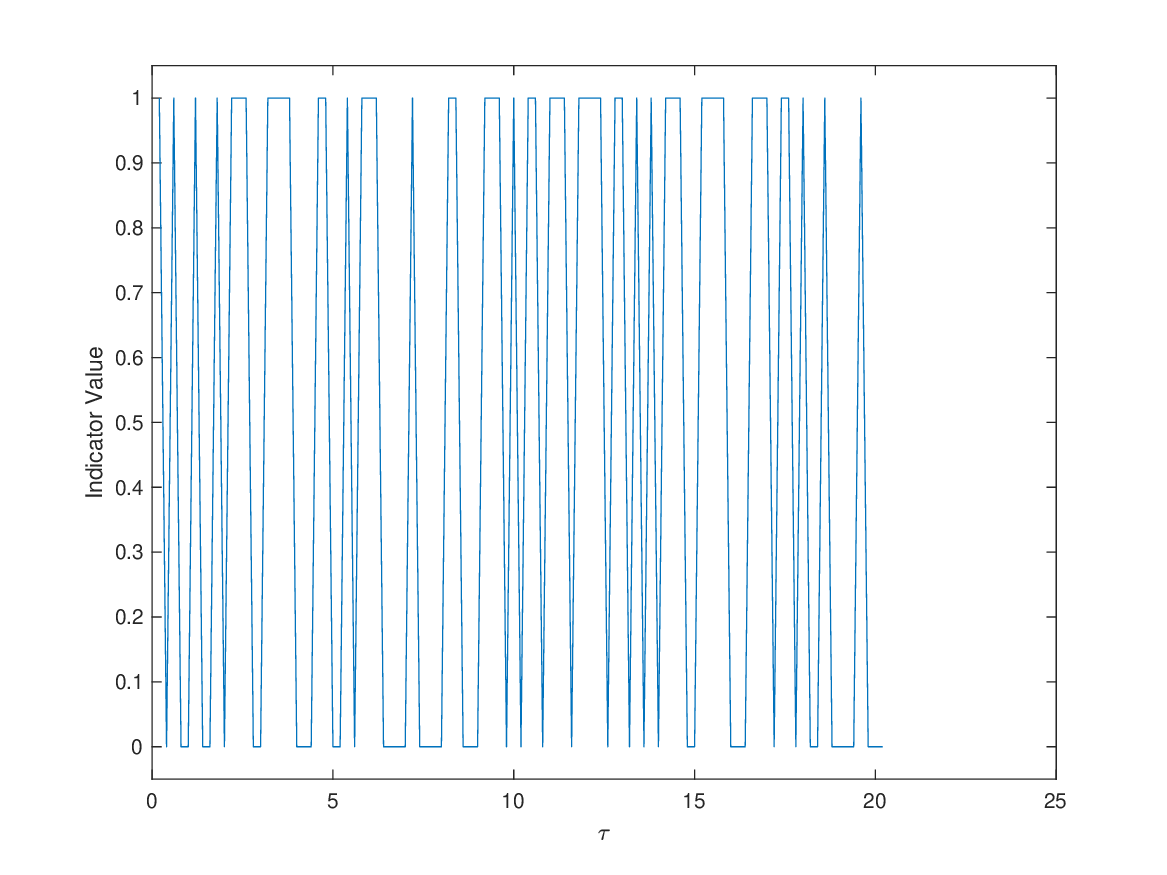}}
\caption{The indicator function}
   \label{fig:fixindicator}
\end{figure}

%  \begin{figure}[ht]
% \centerline{\includegraphics [ height =1.7 in]{discretevaringerror.jpeg}}
% \caption{The trajectory of the norm of the estimation error for systems with noise}
%    \label{fig:dvaryingerror}
% \end{figure}

% With weighted two-norm, $\|\tilde{A}\|_{R(\tau)}=2.30$ where $R(\tau)=\frac{1}{3}I_6$. $\|\tilde{A}\|_{R(\tau)} \|B(\tau)^5\|_{R(\tau)}\leq .5$
% while $ \|\tilde A\|_{R(\tau)}\|B(\tau)^4\|_{R(\tau)}>.5$. Thus by Eq.~\eqref{eq:rnorm},   $q=5$. With mixed matrix norm,
% $\|\tilde{A}\|=2.30$. First choose $p=4$.  $\|\tilde{A}\|\|B^p\|^2\leq 0.5$, thus by Eq.~\eqref{rajit2} $q=2p=8$.

Consider the case when the neighbor graph is switching between Fig.~\ref{graph} (a) and (b) according to Fig.~\ref{fig:fixindicator}.  
 The system considered  has input white noise $v $ which satisfies $v \sim \mathcal N(0,0.1^2)$, that is $x(\tau+1)=Ax(\tau)+v$.
The norm of the estimation error is shown in Fig.~\ref{fig:dvaryingerror} with the same value of $q$.

% Even though the error in Fig.~\ref{fig:dvaryingerror} is larger than the one in Fig.~\ref{fig:di-fix1}, it still does not overshoot the same upper-bound  as that for the fixed neighbor graph case.

% With weighted two-norm for both cases $\|B(\tau)\|_2<1$. Choose $q=6$ so that $\|\tilde{A} B(\tau)^6\|_2\leq 0.5$. With mixed matrix norm, $\|\tilde{A}\|=2.30$. First choose $p=4$, then $\|\tilde{A}\|\|B^p\|^3\leq 0.5 $ for both cases. Thus $q=3p=12$.

 \vspace{-0.1in}
 
\section{Concluding Remarks}
 
 \vspace{-0.1in}

 The distributed state estimation problem is studied   when
the neighbor graph is time-varying. 
It has been shown that, each agent
can estimate the state exponentially fast with a pre-assigned
convergence rate for both continuous-time and discrete-time systems.  

The distributed state estimators developed in both \S \ref{sec:split_continuous} and \S \ref{sec:split_discrete} rely on an especially
 useful observation about distributed estimator structure
  first noted in \cite{Kim2016CDC}  and subsequently exploited in \cite{trent2}  and \cite{Lili19ACC}. 
  That is to split the system spectrum into disjoint subsets corresponding to unobservable and observable subspaces.
  Just how much further this idea
  can be advanced remains to be seen. 
 Generalization on the constraint of strong connectivity for the neighbor graph  can be studied in future work.
Certainly the synchronous switching upon which the local
  estimators in \S \ref{sec:split_discrete}  depend can be relaxed by judicious application of the mixed matrix norm discussed here. This generalization will be addressed in future work.

% There are also  research  \cite{R1,R5} dealing with sensor attacks, where a malicious attacker can manipulate their observations arbitrarily when each sensor only has one dimensional measurement. The study on the resilience and robustness of the algorithms can be further studied.

\begin{appendices}
 \vspace{-0.1in}
\section{Proofs for Continuous-Time Distributed Estimator}
\vspace{-0.1in}
{\bf Proof of Proposition \ref{prop:split_continuous_stable}:} Recall $\bar S=S\otimes I_n$. Since $S$ is a stochastic matrix, $S'$ must have
 a spectral  radius of $1$ and an  eigenvalue  at $1$. Moreover,  since $\mathbb{N}$ is  the graph
 of $S'$ and $\mathbb{N}$  is strongly
 connected by assumption, $S'$ is
  irreducible \{Theorem 6.2.24, \cite{horn}\}. Thus by the Perron-Frobenius Theorem
  there must be a positive vector $\pi$
such that $S'\pi= \pi$. Without loss of generality, assume
  $\pi$  is normalized so that the sum of its entries equals $1$; i.e., $\pi$ is a Perron vector \cite{horn}.
   Let $\Pi $ be that  diagonal matrix
  whose diagonal entries are the entries of $\pi$. Then $\Pi\mathbf{1} = \pi$ where $\mathbf{1}$ is the $m$-vector of all $1$s. Let $L = 2\Pi -\Pi S -S'\Pi$. Clearly
  $L$ is a symmetric matrix and $L\mathbf{1} = 0$.

We claim that the geometric  multiplicity of $L$'s eigenvalue at $0$ is one. To establish this claim, note first that
 every nonzero entry of $S'$ is a nonzero entry of $\Pi S + S'\Pi$
because  $S'$ is a nonnegative matrix and $\Pi$ is a diagonal matrix whose diagonal entries are all positive.
Moreover since $2(I_m-\Pi)$  is a nonnegative matrix,  every nonzero entry of $S'$ is
also a nonzero entry of $2I_m-L = 2(I_m-\Pi) +\Pi S +S'\Pi$. Thus the graph of $S'$ must  be a
  spanning subgraph of the graph of $2I_m-L$ so the graph of $2I_m-L$ must be strongly connected.
  Therefore $2I_m-L$ must be irreducible. Note that the row sums of $2I_m-L$ all equal $2$.
Since   $2I_m-L$ is nonnegative, its infinity norm is $2$ so its spectral radius is no greater than
 $2$ \{Theorem 5.6.9, \cite{horn}\}.
 Moreover $2$ is an eigenvalue of $2I_m-L$.   Thus
by the  Perron-Frobenius Theorem, the geometric multiplicity of this
eigenvalue is one. It follows that the geometric multiplicity of the eigenvalue of $L$ at $0$ is also one.

We claim that $L$ is positive semi-definite. To establish this claim, note that
 $L$ can also be written as $L=D-\widehat{A}$ where $D$ is a diagonal matrix whose entries are the diagonal
   entries of $L$
  and $\widehat{A}$ is the nonnegative matrix  $\widehat{A} =D- L$. As such, $L$ is the generalized Laplacian  \cite{graph}
   of that
  simple undirected graph $\mathbb{G}$  whose adjacency matrix is the matrix which results when the
  nonzero entries $a_{ij}$
  in $\widehat{A}$ are replaced by ones. Since  $L$ can also be written as
   $$L=\sum_{(i,j)\in\mathcal{E}} a_{ij}(u_i-u_j)(u_i-u_j)'$$ 
where $u_i$ is the $i$th unit vector and $\mathcal{E}$ is the edge
   set of $\mathbb{G}$, $L$ is
    positive semi-definite as claimed.

To proceed, set
\begin{equation}\label{eq:H} H =\text{ block diag }\{\pi_1I_{n_1}, \pi_2I_{n_2},\ldots, \pi_{m}I_{n_m}\}\end{equation} where $n_i = \dim \mathcal{V}_i$
 and note that
$VH = (\Pi\otimes I_n)V$.  Since $((S-I_m)'\otimes I_n)(\Pi\otimes I_n) =((S-I_m)'\Pi)\otimes I_n$
 it must be true that
$(V'((S-I_m)\otimes I_n)V)'H = V' (((S-I_m)'\Pi)\otimes I_n)V$
and thus that
\begin{align}
&\; H(V'(I_{mn}-\bar S)V)+ (V'(I_{mn}-\bar S)V)'H  \nonumber\\ 
&= V'(L\otimes I_n)V\label{eq:split_continuous_lyap}\end{align}
Observe that this is a Lyapunov equation for the positive definite function $z'Hz$.
Therefore to show that  $-V'(I_{mn}-\bar S)V$  is a stability matrix, it is enough to show that
$V'(L\otimes I_n)V$ is positive definite.

Since $L$ is positive semi-definite, so must be $L\otimes I_n$. Therefore $V'(L\otimes I_n)V$ is at
least positive semi-definite.
Suppose $z'V'(L\otimes I_n)Vz = 0$ where $z = $ col $\{z_1,z_2,\ldots,z_m\}$ and
  $z_i\in\R^{\dim(\mathcal{V}_i)}$. To show that $V'(L\otimes I_n)V$ is positive definite,
 it is enough to show that   $z=0$. Since $L$'s eigenvalue at $0$ has multiplicity one,
  $\ker L = $ span $\{\mathbf{1}\}$; therefore  $\ker (L\otimes I_n) = $ column span
 $\mathbf{1}\otimes I_n$. The hypothesis $z'V'(L\otimes I_n)Vz = 0$ implies that $(L\otimes I_n)Vz = 0$
 so $Vz\in \ker (L\otimes I_n)$. Therefore
 $V_iz_i=V_jz_j,\;i,j\in\mathbf{m}$. But because of joint observability,
$\bigcap_{i\in\mathbf{m}} \mathcal{V}_i = 0$ so $V_iz_i = 0,\;i\in\mathbf{m}$. Thus $z_i = 0,\;i\in\mathbf{m}$
 so $z=0$  implying that $V'(L\otimes I_n)V$ is positive definite.  Therefore $-V'(I_{mn}-\bar S)V$
  is a continuous-time stability matrix as claimed.
\hfill  $\qed $

{\bf Proof of Theorem \ref{thm:split_continuous_constant}:}
Recall that the state estimation error satisfied \eqref{eq:split_continuous_err}. The overall error dynamic is defined as \eqref{eq:split_continuous_em}.
According to \eqref{eq:split_continuous_split}, the spectrum of  $\bar{A} -gP(I_{mn}-\bar S)$ is equivalent to the union of the spectrum of $\bar A_V$ and $A_V$.
Since the spectrum of  $\bar{A}_i+\bar{K}_i\bar{C}_i,\;i\in\mathbf{m}$, is assignable with  $\bar{K}_i$, to show for $g$ sufficiently large that $\bar{A} -gP(I_{mn}-\bar S)$ is a continuous-time stability matrix with a prescribed convergence rate  as large as $\lambda$, it is enough to show that for $g$ sufficiently large, the matrix $A_V=\tilde{A}-gV'(I_{mn}-\bar S)V$ is a continuous-time stability matrix with a prescribed convergence rate as large as $\lambda$.

To show that
$\text{exp}\{(\tilde{A} - gV'((I_m-S)\otimes I_n)V)t\}$
 can be made to converge to zero as fast as $\text{exp}({-\lambda t})$ does
  by choosing  $g$  sufficiently large,
   we exploit \eqref{eq:split_continuous_lyap}. Note in particular that
%$$H(\lambda I+ \tilde{A} - gV'(I-S)\otimes I)V) + (\lambda I+ \tilde{A} - gV'(I-S)\otimes I)V)'H =
%H(\lambda I+ \tilde{A}) + (\lambda I+ \tilde{A})'H - gV'(L\otimes I)V$$
\begin{eqnarray*}
&\ &H(\lambda I+ A_V) + (\lambda I+  A_V)'H\\& =&
H(\lambda I+ \tilde{A}) + (\lambda I+ \tilde{A})'H - gV'(L\otimes I_n)V
\end{eqnarray*}
Since   $V'(L\otimes I_n)V$ is positive definite, by picking $g$ sufficiently large,
$H(\lambda I+ \tilde{A}) + (\lambda I+ \tilde{A})'H - gV'(L\otimes I_n)V$ will be negative definite implying that
%$\lambda I+ \tilde{A} - gV'(I-S)\otimes I)V$
 $\lambda I+  A_V$ is a stability matrix and thus that $\tilde{A} - gV'(I_{mn}-\bar S)V$
 is a stability matrix for which $\text{exp}\{(\tilde{A} - gV'(I_{mn}-\bar S)V)t\}$
  converges to zero as fast as $\text{exp}({-\lambda t})$ does. In other words,
  any value of $g$ will have the desired property provided  
  \begin{equation}\label{eq:g} g\geq \frac{\lambda_{\max}\left(H(\lambda I+ \tilde{A}) + (\lambda I+ \tilde{A})'H\right)}{\lambda_{\min}\left(V'(L\otimes I_n)V\right)}\end{equation} where $\lambda_{\max} (.) $ and  $\lambda_{\min}(.)$ 
 are the largest eigenvalue and the smallest eigenvalue of a symmetric  matrix respectively.
 \hfill  $\qed $

\noindent \textbf{Proof of Lemma~\ref{lem:split_continuous_average}:} \footnote{The symbols used in this proof such as $g$, $c$ and $\lambda^*$ are generic and do
not have the same meanings as the same symbols do when used elsewhere in
this paper.}
By hypothesis, each
 $M_i$  is exponentially  stable. Thus there  are positive constants $c_i>1$ and $\lambda_i$ such that
\begin{equation}
\|\text{exp}({M_it})\|\leq c_i \text{exp}({-\lambda_i t})   \label{eq:split_continuous_lemma2} 
\end{equation}
for any $i\in \{1,2,\ldots,|\mathcal{G}|\}$.
Here  $\| \cdot \|$ is any given submultiplicative norm on $\mathbb{R}^{n\times n}$.
 Let 
 $$
c=\max_{i\in \{1,2,\ldots, |\mathcal{G}|\}} c_i,\;\text{and }
\lambda^*=\min \limits_{i\in\{1,2,\ldots, |\mathcal{G}|\}} \lambda_i. $$

Fix $\lambda>0$ and let $g$ be any gain satisfying\eq{g\geq \frac{\tau_D(\lambda +\|N\|c) +\ln c}{\tau_D\lambda^*}\label{eq:split_continuous_lili}}  
We claim that for any number $\tau$, and any switching signal $\sigma\in \mathcal{S}_{\text{avg}}$,  the transition matrix of $gM_\sigma$, namely $\Phi_{\sigma}(t,\tau)$, converges to zero as fast as $\text{exp}({-\alpha t})$ does where %$\alpha$ is defined in \eqref{eq:split_continuous_billi}.
\eq{\alpha = \lambda +\|N\|c\label{eq:split_continuous_billi}}
To understand why this is so, by \eqref{eq:split_continuous_lemma2},
\begin{equation}\label{eq:split_continuous_phi}
\|\Phi_{\sigma}(t,\tau)\|\leq  c^{\delta_{\sigma}(\tau,t)}\text{exp}({-g\lambda^*(t-\tau)})
\end{equation}
where $\delta_{\sigma}(\tau,t)$  is the number of switching between $(\tau,t)$.
%and $\delta_{\sigma}(\tau,t)\leq \delta_0+\frac{t-\tau}{\tau_D}  $.
By \eqref{eq:split_continuous_lili}, $\text{exp}({g\lambda^*})\geq c^{\frac{1}{\tau_D}}\text{exp}({\alpha})$. 
From this and the fact that  $\delta_{\sigma}(\tau,t)\leq \delta_0+\frac{t-\tau}{\tau_D}  $,
\begin{multline*}
    \|\Phi_{\sigma}(t,\tau)\|\leq  c^{\delta_0}c^{\frac{t-\tau}{\tau_D}}\text{exp}({-g\lambda^*(t-\tau)})\\ \leq c^{\delta_0-\frac{\tau}{\tau_D}}\text{exp}({-\alpha (t-\tau)}).
\end{multline*}
 
Thus, the claim is true.

% The variation of constants formula for \eqref{eq:split_continuous_sneeze} is \eqref{eq:split_continuous_x} where $\Phi_{\sigma(t,\tau)}$ is the transition matrix of $gM_\sigma$ for each $\sigma\in \mathcal{S}_{\text{avg}}$.

In view of \rep{eq:split_continuous_sneeze} and the variation of constants formula,
\begin{equation}\label{eq:split_continuous_x}
x(t)=\Phi_{\sigma}(t,0) x(0)+\int_0^t\Phi_{\sigma}(t,\mu)Nx(\mu)d\mu
\end{equation}

As
$\|\Phi_{\sigma}(t,\tau)\|\leq c\ \text{exp}({- \alpha t})$ for all  $\tau$ and  $\sigma\in\mathcal{S}_{\text{avg}}$,
\begin{multline*}
    \|x(t)\|  \leq   c\  \text{exp}({-\alpha t}) \|x(0)\| +\\ \int_0^tc\  \text{exp}({-\alpha (t-\mu)} ) \|N\|\|x(\mu)\|d\mu
\end{multline*}
% which is the same as \eqref{eq:split_continuous_x_inequality}. The rest proof to show 
% $$\|x(t)\|\leq c \|x(0)\| e^{-\lambda t}$$
% is exactly the same as the proof of Lemma~\ref{lem:split_continuous_dwell} which is omitted here. \hfill
By multiplying by $\text{exp}({\alpha t})$ on both sides, one obtains
$\text{exp}({\alpha t} )\|x(t)\|  \leq   c \|x(0)\| +\int_0^t ||N||c\  \text{exp}({\alpha \mu}) \|x(\mu)\|d\mu$.
From this and the Bellman-Gronwall Lemma there follows
\[\text{exp}({\alpha t})\|x(t)\|  \leq  c \|x(0)\| \ \text{exp}({\int_0^t \|N\|c d\mu })
\]
Since $\int_0^t \|N\|c d\mu=\|N\|ct$, it follows that $\text{exp}({\alpha t}) \|x(t)\|  \leq c\|x(0)\| \text{exp}({\|N\|ct })$ and thus that 
\[\|x(t)\|\leq c \|x(0)\| \text{exp}({(\|N\|c-\alpha)t })
\]
From this and \rep{eq:split_continuous_billi} it follows that 
$$\|x(t)\|\leq c \|x(0)\| \text{exp}({-\lambda t})$$
This completes the proof.\hfill  
$\qed$

 \vspace{-0.1in}
 
\noindent {\bf Proof of Theorem~\ref{thm:split_continuous_average}:} 
Recall that $A_V(t)=\tilde{A}-gV'(I_{mn}-\bar S(t))V$.
 By Proposition~\ref{prop:split_continuous_stable}, for any fixed time $\tau$, $-V'(I_{mn}-\bar S(\tau))V$ is exponentially stable if the graph of $S(\tau)'$ is strongly connected.
 Note $\tilde{A}$ is fixed and bounded.
According to Lemma~\ref{lem:split_continuous_average}, for each $\sigma\in \mathcal{S}_{\text{avg}}$ there is a positive number $g$, depending on $\tau_D$ so that the transition matrix of $A_V(t)$ converges to zero at least as fast as $\text{exp}({-\lambda t})$   does.
This is accomplished by choosing $g$ sufficiently large.
 Based on the proof of Lemma~\ref{lem:split_continuous_average}, it is sufficient to pick $g$ to satisfy
  
 \vspace{-0.1in}
 
\begin{equation}\label{eq:split_continuous_g}
    g\geq \frac{\ln c +(\lambda +\|\tilde{A}\|c)\tau_D}{\lambda^* \tau_D}
\end{equation} 
 
 \vspace{-0.1in}
where  $c$ and $\lambda^*$ are two positive numbers chosen so that for any fixed $\tau$,
$\|\text{exp}\{-V'(I_{mn}-\bar S(\tau))Vt\}\|\leq c \  \text{exp}({-\lambda^* t})$,
and $c>1$. This completes the proof. \hfill$\qed$

\noindent {\bf Proof of Theorem~\ref{thm:arb}:} 
Recall $\Phi_{V}(t,\tau)$ is the transition matrix of $A_V(t)$ for any $t\geq \tau \geq 0$.
If we can show that there exist a constant $c$ so that
\[\|\Phi_V(t,\tau)\|\leq c\  \text{exp}({-\lambda(t-\tau)}),\;\;\; \forall t\geq \tau\geq 0\]
the remaining proof is exactly the same as the proof of Theorem~\ref{thm:split_continuous_constant} which is omitted here. 

It is left to show that $\|\Phi_V(t,\tau)\|\leq c \ \text{exp}({-\lambda(t-\tau)})$ for all $t\geq \tau\geq 0$ by choosing $g$ sufficiently large.
We explore the matrix $A_V(t)$.
Recall that $A_V(t)=\tilde{A}-gV'((I_m-S(t))\otimes I_n)V$. 
In particular, 
\begin{eqnarray*}
&\ & (\lambda I + A_V(t))+(\lambda I + A_V(t))'\\ &=&
(\lambda I+\Tilde{A})+(\lambda I+\Tilde{A})'\\ &\ &-gV'((2I_m-S(t)-S'(t))\otimes I_n)V
\end{eqnarray*}
Since each $S(t)$ is doubly stochastic, $2I_m-S(t)-S'(t)$ has row sum $0$,  all its off-diagonal entries are non-positive, and all its diagonal entries are positive. 
That is, this matrix can be seen as a generalized Laplacian matrix of a connected graph.
By Proposition~\ref{prop:split_continuous_stable}, for any $t$,
$-V'((2I_m-S(t)-S'(t))\otimes I_n)V = -V'(I_{mn}-\bar S(t))V  -V'(I_{mn}-\bar S'(t))V$ is negative definite.
Thus by picking $g$ sufficiently large,  $ (\lambda I + A_V(t))+(\lambda I + A_V(t))'$ will be negative definite for any time $t$.

Consider system 
\[\dot {\bar z}=A_V(t) \bar z\]
Let $V=\bar z'\bar z$. Then
\[
\dot{V}=\bar z'(A_V(t)'+A_V(t))\bar z\leq -2\lambda \bar z'\bar z
\]
Therefore, $\Phi_V(t,\tau)$  converges to zero as fast as $\text{exp}({-\lambda (t-\tau)})$ does, i.e.,
\[\|\Phi_V(t,\tau)\|\leq c\ \text{exp}({-\lambda(t-\tau)}),\;\;\; \forall t\geq \tau\geq 0\]
This completes the proof. 
\hfill$\qed$
 
\noindent {\bf Proof of Theorem~\ref{thm:split_continuous_adaptive}:} 
Equation~\eqref{eq:adaptive} can be
rewritten as \begin{equation}\label{eq:gain} \dot
g_i=|V_i'W_ie|_2^2,\;\;\;i\in\mathbf m
\end{equation}
where $W_i=\begin{bmatrix}W_{i1}& \ldots & W_{im}\end{bmatrix} \in \mathbb R^{n\times nm}$.
Here $ W_{ij}\in \mathbb{R}^{n\times n}$ is $\frac{1}{m_i} I_n $ if $j\neq i$ and $j\in \mathcal{N}_i$, and $W_{ij}$ is  $-I_n$ if $j=i$. Otherwise $W_{ij}$ is a $0$ matrix.  
Let
$\text{column}\{W_1,W_2,\ldots, W_m\}=W$.

Different from \eqref{eq:split_continuous_em}, the error model
 turns to
 \begin{equation} \label{eq:split_cotinuous_emnew}
     \dot{e}=(\bar A-G(t)P(I_{mn}-\bar S(t))e
 \end{equation}
 where $G(t)=\text{ block diag }\{g_1(t) I_n,\ldots, g_m(t) I_n\}$.
 Let $\begin{bmatrix}
 (Qe)' & z'
 \end{bmatrix}' =T^{-1}e$ where $T=\begin{bmatrix}
 Q^{-1} &V
 \end{bmatrix}$ as defined earlier. Here $z=[z_1',\ \ldots,z_m']'$ iwth $z_i=V_i'e_i$.
 
Based on \eqref{eq:split_continuous_split} and \eqref{eq:split_cotinuous_emnew}, the dynamic of $z_i$ can be
written in the following form
% \begin{equation}\label{eq:znew}
% \dot z=\tilde Az -\bar GV'WVz+\hat AQe-\bar GV'WQ'Qe
% \end{equation}
% Correspondingly,
\begin{equation}\label{eq:zi}
\dot z_i=\tilde A_iz_i -g_i(t)V_i'M_ie+\hat A_iQ_ie_i,\;\;i\in \mathbf
m
\end{equation}
where $\tilde A_i=V_i'(A+K_iC_i)V_i$, and $\hat
A_i=V_i'(A+K_iC_i)Q_i'$.

First, we want to show that all $g_i(t)$ are bounded. We prove this by
contradiction.
Without generality, suppose that $g_i$ for $i\in
\mathcal V_u=\{1,2,\ldots, m_1\}$ are unbounded, and $g_i$ for $i\in
\mathcal V_b=\{m_1+1,m_1+2,\ldots, m\} $ are bounded where $\mathcal
V_u\cap \mathcal V_b=0$ and $\mathcal V_u\cup \mathcal V_b=\mathbf
m$.

Let $R=R_1+R_2+R_3+R_4$ where the individual $R_i$ involve new positive parameters $p$, $\alpha_0$, $\alpha_{m_1+1},\dots, \alpha_m$.
\[R_1=\frac{1}{2} \sum_{i=1}^{m_1}
\pi_i \frac{p}{g_i(t)} |z_i|_2^2,\;\; R_2=\frac{1}{2}\sum_{i=m_1+1}^m
\pi_i \frac{g_i(0)}{g_i(t)}|z_i|_2^2\]
\[R_3=-\sum_{i=m_1+1}^{m}\alpha_ig_i(t), \text{and } R_4=-\alpha_0\int_{0}^t|Qe|_2^2dt.\]  
The way to pick positive parameters $p$, $ \alpha_0$, and $\alpha_i$
for $i\in\mathcal V_b$ is specified as follows.

{\bf Picking $p\geq 1$:}

Let $\mathcal{W}_1$ be a positive matrix matrix chosen such that
\begin{equation}\label{eq:D} z'\mathcal{W}_1z=\sum_{i=1}^m \pi_i|\tilde A_i|_2 |z_i|_2^2
\end{equation}
According to \eqref{eq:split_continuous_lyap},
\begin{equation}\label{eq:R}
  F=\frac{1}{2}(HV'(\bar S-I_{mn})V+V'(\bar S-I_{mn})'VH)>0
\end{equation}
 Pick $p$   so that
$ \mathcal W_2=pF-\mathcal W_1>0
$

{\bf Picking $\alpha_0$, and $\alpha_i$, $i\in \mathcal V_b$:}

Using the Cauchy-Schwarz inequality, the following three inequalities
can be derived. For $\beta_1 >0, \; \beta_2>0$ and $\lambda_i
>0{\text{ for }} i\in \mathcal V_b$, all for the moment otherwise arbitrary, write $\hat A=\text{block diag }\{\hat A_1,\ldots,\hat A_m\}$
\vspace{-.2in}
{\small \begin{equation}
\label{eq:beta1}
 z'H\hat A Qe  \leq  \frac{\beta_1}{2}|\hat A'
    Hz|_2^2+\frac{1}{2\beta_1}|Qe|_2^2,
\end{equation}
\vspace{-.3in}
\begin{multline}
  -p z'HV'(\bar S-I_{mn})Q' Qe  \leq   \\  \frac{\beta_2}{2}|Q(\bar S-I_{mn})'V
    Hz|_2^2+\frac{1}{2\beta_2}|Qe|_2^2\label{eq:beta2}
\end{multline}
\vspace{-.3in}
\begin{multline}
 - \pi_i (g_i(0)-p) z_i'V_i'W_ie \leq\\
\frac{\lambda_i}{2}|\pi_i (g_i(0)-p) z_i|_2^2 
 +\frac{1}{2\lambda_i}|V_i'W_ie|_2^2.\label{eq:sum}
\end{multline}
\vspace{-.2in}
Let   $\mathcal W_3$ be a symmetric  matrix chosen such that
\begin{eqnarray}
    z'\mathcal W_3z&=&z'\mathcal W_2z-\frac{\beta_1}{2}|\hat A'Hz|_2^2-\frac{\beta_2}{2}|Q(\bar S-I_{mn})'VHz|_2^2 \nonumber\\
    &\ &-\sum_{i=m_1+1}^m \frac{\lambda_i}{2} |\pi_i(g_i(0)-p)z_i|_2^2\label{eq:finalL}
\end{eqnarray}}
Pick positive $\beta_1$, $\beta_2$, and  $\lambda_i$  for $i\in \mathcal
V_b$ to be small enough so that $W_3$ is positive definite. Pick
$\alpha_0$ and $\alpha_i$, $i\in \mathcal V_b$ according to the following two equations:
\vspace{-.2in}
{\small \begin{equation}
\alpha_0=\frac{1}{2\beta_1}+\frac{1}{2\beta_2}
\;\;\text{and }\;\;
    \alpha_i=\frac{1}{2\lambda_i} \;\;\;\;\; i\in \mathcal{V}_b \label{eq:alpha}
\end{equation}}
% \eqref{eq:alpha0}, and \eqref{eq:alpha}.
% \begin{equation}\label{eq:alpha0}
% \alpha_0=\frac{1}{2\beta_1}+\frac{1}{2\beta_2}
% \end{equation}
% \begin{equation}\label{eq:alpha}
%     \alpha_i=\frac{1}{2\lambda_i} \;\;\;\;\; i\in \mathcal{V}_b
% \end{equation}

\vspace{-.2in}
Now we consider the derivative of $R$, i.e., $ \dot R=\dot R_1+\dot
R_2+\dot R_3+\dot R_4$.
\vspace{-.2in}
{\small \begin{eqnarray*}
\dot R_1&=&-\frac{1}{2}\sum_{i=1}^{m_1} \pi_i \frac{p}{g_i^2}
|z_i|_2^2|V_i'W_ie|_2^2+ \sum_{i=1}^{m_1} \pi_i \frac{p}{g_i }
z_i'\tilde A_iz_i \nonumber \\ &\ &+ \sum_{i=1}^{m_1} \pi_i
\frac{p}{g_i } z_i'\hat A_iQ_ie_i-\sum_{i=1}^{m_1} \pi_i p
z_i'V_i'W_ie
\end{eqnarray*}}
Since for $i\in \mathcal V_u$  each $g_i(t)$ is   unbounded, there is a time
$T$ for $t\geq T$, $p\leq g_i(t)$ for all $i \in \mathcal V_u$. That is $\frac{p}{g_i(t) }\leq
1$. Hence
\vspace{-0.4in}
{\small \begin{eqnarray}\label{eq:R1}
\dot R_1\leq   \sum_{i=1}^{m_1} \pi_i  z_i'\tilde A_iz_i +
\sum_{i=1}^{m_1} \pi_i z_i'\hat A_iQ_ie_i-\sum_{i=1}^{m_1} \pi_i p
z_i'V_i'W_ie
\end{eqnarray}}
According to \eqref{eq:gain}, $g_i(t)$ is a non-decreasing function. As
a result, $\frac{g_i(0)}{g_i(t)}    \leq  1  $. Similar to the derivation
of the inequality for $\dot R_1$, it can be shown that
\vspace{-.2in}
{\small \begin{equation}
\dot R_2  \!\!\leq \!\! \!\!\!\!  \sum_{i=m_1+1}^{m} \!\!\!\!\! \!\pi_i
z_i'\tilde A_iz_i +
\!\!\!\!\!  \sum_{i=m_1+1}^{m} \! \!\!\!\!  \pi_i z_i'\hat A_iQ_ie_i  - \! \!\!\!\! \sum_{i=m_1+1}^{m} \! \!\!\!\!  \pi_i
 g_i(0)
z_i'V_i'W_ie  
\label{eq:R2}\end{equation}}
By the submultiplicity of the matrix two norm, $z_i'\tilde A_iz_i'
\leq   |\tilde A_i|_2 |z_i|_2^2$. From this, \eqref{eq:R1},
\eqref{eq:R2}, \eqref{eq:D} and \eqref{eq:H},
\vspace{-0.2in}
{\small
\begin{eqnarray*}
\dot R_1+\dot R_2&\leq& z'\mathcal W_1z+\sum_{i=1}^{m} \pi_i z_i'\hat
A_iQ_ie_i-\sum_{i=1}^{m} \pi_i p z_i'V_i'W_ie  \nonumber\\
&\ & -\sum_{i=m_1+1}^{m} \pi_i (g_i(0)-p) z_i'V_i'W_ie \nonumber\\
&=&z'\mathcal W_1z+z'H\hat A Qe-pz'HV'(\bar S-I_{mn})e\\
&\ &-\sum_{i=m_1+1}^{m} \pi_i (g_i(0)-p) z_i'V_i'W_ie
\end{eqnarray*}}
 \vspace{-0.2in}
 
It can be observed that
$z'HV'(\bar S-I_{mn})e=z'HV'(\bar S-I_{mn})(VV'e+Q'Qe)=z'HV'(\bar S-I_{mn})Vz+z'HV'(\bar S-I_{mn})Q'Qe$. From this and
\eqref{eq:R}, $z'HV'((S-I_m)\otimes I_n)e=z'Fz+z'HV'((S-I_m)\otimes I_n)Q'Qe$. Thus
\vspace{-0.2in}
{\small \begin{multline*}
\dot R_1+\dot R_2\leq  z'(\mathcal W_1-pF)z+z'H\hat A Qe- pz'HV'\\(\bar S-I_{mn})Q'Qe
-\sum_{i=m_1+1}^{m} \pi_i (g_i(0)-p) z_i'V_i'W_ie \\
= -z'\mathcal W_2z+z'H\hat A Qe-pz'HV'\\(\bar S\!-\!I_{mn}\!)Q'Qe-\sum_{i=m_1+1}^{m} \pi_i (g_i(0)-p) z_i'V_i'W_ie
\end{multline*}}
\vspace{=-0.3in}
From this, \eqref{eq:beta1}, \eqref{eq:beta2}, and \eqref{eq:sum},
{\small \begin{eqnarray}\nonumber
\dot R_1+\dot R_2&\leq& -z'\mathcal W_2z+\frac{\beta_1}{2}|\hat A'Hz|_2^2+\frac{\beta_2}{2}|Q(\bar S-I_{mn})'VHz|_2^2 \nonumber\\
    &\ &+\sum_{i=m_1+1}^m \frac{\lambda_i}{2}
    |\pi_i(g_i(0)-p)z_i|_2^2+\frac{1}{2\beta_1}|Qe|_2^2 \nonumber\\
    &\ & +\frac{1}{2\beta_2}  |Qe|_2^2+\sum_{i=m_1+1}^m
    \frac{1}{2\lambda_i}|V_i'W_ie|_2^2\label{eq:R12}\end{eqnarray}}
 It is direct to get that
{\small \begin{equation}\label{eq:R34}\dot R_3+\dot
R_4=-\sum_{i=m_1+1}^{m}\alpha_i|V_i'W_ie|_2^2-\alpha_0 |Qe|_2^2
\end{equation}}
From \eqref{eq:R12}, \eqref{eq:R34}, and
\eqref{eq:alpha},
\vspace{-0.2in}
{\small \begin{eqnarray*}
\dot R&\leq& -z'\mathcal W_2z+\frac{\beta_1}{2}|\hat A'Hz|_2^2+\frac{\beta_2}{2}|Q((S-I)\otimes I)'VHz|_2^2 \nonumber\\
    &\ &+\sum_{i=m_1+1}^m \frac{\lambda_i}{2}
    |\pi_i(g_i(0)-p)z_i|_2^2
\end{eqnarray*}}
According to \eqref{eq:finalL},
\begin{equation}\label{eq:finalfinal}
\dot R\leq -z'\mathcal W_3z<0\end{equation}
Since $Qe$ is exponentially convergent, the limit of $R_4(t)$ as $t$
goes to infinity exists. Due to the assumption that for $i\in
\mathcal V_b$, the $g_i$ are bounded. Thus $R_3$ is bounded. Therefore,
$R$ is lower bounded. From this and \eqref{eq:finalfinal} it follows
that $z\in \mathcal L^2$. From this and the fact that $Qe\in
\mathcal L^2$, we conclude $e\in \mathcal L^2$.  This   with the definition of
$\dot g_i$ imply that all $g_i$ for $i\in \mathbf{m}$ are bounded.
Thus by contradiction all $g_i$ are bounded.

Next, we want to show that $e$ converges to zero. According to
Theorem  \ref{thm:split_continuous_constant}, let $G_1=g_1I$ be a matrix chosen so that $\bar
A-G_1P(I_{mn}-\bar S(t))$ is a stable matrix. From \eqref{eq:split_cotinuous_emnew}, the dynamic of $e$ can
be rewritten as
\begin{equation} \label{eq:final2022}
\dot e=(\bar A-G_1P(I_{mn}-\bar S(t)))e+(G_1-G)P(I_{mn}-\bar S(t))e\end{equation} Since $(I_{mn}-\bar S(t))e\in
\mathcal L^2$, $G_1-G$ is bounded, and $\bar A-G_1P(I_{mn}-\bar S(t))$ is stable, thus $\dot e\in \mathcal L^2$.
Thus \eqref{eq:final2022} is  input-to-state stable which implies that $e$  must converge to zero asymptotically.
\hfill
$\qed$

 \section{Proofs for Discrete-Time Distributed Estimator}
\vspace{-0.1in}
\noindent{\bf Proof of Lemma \ref{lem:split_discrete_brian}:} 
%Since $M$ is a stochastic
% matrix, it must have
%  a spectral  radius of $1$ and an  eigenvalue  at $1$ as must $M'$.  Moreover,  since
%  the graph of $M'$ is strongly
%  connected, $M'$ is
%   irreducible \{Theorem 6.2.24, \cite{horn}\}. Thus by the Perron-Frobenius Theorem there
%    must be a positive vector $\pi$
% such that $M'\pi= \pi$. Without loss of generality, assume
%   $\pi$  is normalized so that the sum of its entries equals $1$; i.e., $\pi$ is a probability vector.
%    Let $\Pi_M $ be that  diagonal matrix
%   whose diagonal entries are the entries of $\pi$. Then $\Pi_M\mathbf{1} = \pi$.
Since $M$ is an $m\times m$ row stochastic matrix which ahs a strongly connected graph, $M$   is
    irreducible \{Theorem 6.2.24, \cite{horn}\}. Thus by the Perron-Frobenius Theorem there
   must be a positive vector $\pi$
such that $M'\pi= \pi$.
Without loss of generality, assume
  $\pi$  is normalized so that the sum of its entries equals $1$; i.e., $\pi$ is a probability vector.
   Let $\Pi_M $ be that  diagonal matrix
  whose diagonal entries are the entries of $\pi$. Then $\Pi_M\mathbf{1} = \pi$.
 Since  $M\mathbf{1} = \mathbf{1}$,  $\Pi_M \mathbf{1} = \pi$, and
$M'\pi = \pi$,   it must be true that $M'\Pi_M M\mathbf{1} = \pi$
and thus that $L_M\mathbf{1} = 0$. 
Thus $L_M$ can also be written as $L_M=D-\hat{A}$ where $D$ is a
  diagonal matrix whose diagonal entries are the diagonal
  entries of $L_M$
  and $\hat{A}$ is the nonnegative matrix $\hat{A} = D- L_M$.
Arguing as in the proof of Proposition~\ref{prop:split_continuous_stable}, it can be shown that $L_M$ is
 positive-semidefinite.
% To show that $L_M$ is
% positive-semidefinite note first that
%  $L_M$ can also be written as $L_M=D-\hat{A}$ where $D$ is a
%   diagonal matrix whose diagonal entries are the diagonal
%   entries of $L_M$
%   and $\hat{A}$ is the nonnegative matrix $\hat{A} = D- L_M$.
%   As such, $L_M$ is the generalized Laplacian  \cite{graph}
%   of that
%   simple undirected graph $\mathbb{G}$  whose adjacency matrix is the matrix which results when the
%   nonzero entries $a_{ij}$
%   in $\hat{A}$ are replaced by ones. Since  $L_M$ can also be written as
%   $$L_M=\sum_{(i,j)\in\mathcal{E}} a_{ij}(u_i-u_j)(u_i-u_j)'$$ where $u_i$ is the $i$th unit vector
%     and $\mathcal{E}$ is the edge
%   set of $\mathbb{G}$, $L_M$ is
%     positive semi-definite as claimed.

Now suppose that the diagonal entries of $M$ are all positive. Then
the diagonal entries  of $M'\Pi_M $ must also all be positive. It
follows that
 every arc in the graph of $M'$ must be an arc in the graph of $M'\Pi_M M$ so the graph of $M'\Pi_M M$ must
  be strongly connected. Since $I-\Pi_M$ is a nonnegative matrix, the graph of $M'\Pi_M M$ must be
 a spanning subgraph
   of the graph of $I-\Pi_M +M'\Pi_M M$.
 Since $I-L_M = I-\Pi_M +M'\Pi_M M$ and the graph of $M'\Pi M$ is strongly connected, the graph of $I-L_M$ must
  be strongly connected as well.
But $I-L_M$ is a nonnegative matrix so it must be irreducible. In
addition, since
   $(I-L_M)\mathbf{1} = \mathbf{1}$, the row sums of $(I-L_M)$ all equal one. Therefore the infinity norm of
    $I-L_M$ is one so its spectral radius is no greater than $1$. Moreover $1$ is an eigenvalue of $I-L_M$.
     Thus
by the  Perron-Frobenius Theorem, the geometric multiplicity of this
eigenvalue is one. It follows that the geometric multiplicity of the
eigenvalue of $L_M$ at $0$ is also one;
 ie, the dimension of the kernel of $L_M$ is one as claimed. \hfill \qed

\noindent{\bf Proof of Proposition \ref{prop:split_discrete_mp}:} Fix $\tau $ and write
$S$ for $S(\tau)$ and $\bar{S}$ for $\bar{S}(\tau)$. Note that the
graph of $S'$, namely $\mathbb{N}$,   is strongly connected.
 In view of Lemma \ref{lem:split_discrete_brian}, the matrix $L = \Pi_S-S'\Pi_S S$
 is positive semi-definite and $L\mathbf{1} = 0$. Moreover, since
the diagonal entries of $S$ and thus $S'$  are all positive, the
kernel of $L$ is one-dimensional.

Write $R$ for $R(\tau )$.    %$R =  V'\bar{\Pi} V $ where  $\bar{\Pi} = \Pi(\tau) \otimes I_n$.
To prove the proposition  it is enough to show that the matrix \eq{
Q = R- (V'\bar{S}'V)R(V'\bar{S}V)\label{eq:split_discrete_1}} is positive definite.

\vspace{-0.1in}

To proceed, set   $\bar{L} = L\otimes I_n$ in
 which case $\bar{L}$ is positive semi-definite because $L$ is. Moreover,
 $\bar{L} = \bar{\Pi}- \bar{S}'\bar{\Pi }\bar{S}$ where $\bar{\Pi} = \Pi_{S}\otimes I_n$.
Note that $VRV'  = P\bar{\Pi}P$ where $P$ is the orthogonal
projection matrix $P=VV'$. Clearly $VRV' =
P\bar{\Pi}^{\frac{1}{2}}\bar{\Pi}^{\frac{1}{2}}P$. Note that both
$P$ and $\bar{\Pi}^{\frac{1}{2}}$ are block diagonal matrices with
corresponding diagonal blocks of the same size. Because of this and
the
 fact that each diagonal block in $\bar{\Pi}^{\frac{1}{2}}$
 is a scalar times the identity matrix, it must be true that $P$ and $\bar{\Pi}^{\frac{1}{2}}$ commute; thus
$P\bar{\Pi}^{\frac{1}{2}} = \bar{\Pi}^{\frac{1}{2}}P$. From this and
the fact that $P$ is
 idempotent, it follows that
$VRV' =\bar{\Pi}^{\frac{1}{2}}P\bar{\Pi}^{\frac{1}{2}}$.
 Clearly $\bar{\Pi}^{\frac{1}{2}}P\bar{\Pi}^{\frac{1}{2}}\leq
\bar{\Pi}^{\frac{1}{2}}\bar{\Pi}^{\frac{1}{2}}$ so $VRV' \leq
\bar{\Pi}$. It follows using \eqref{eq:split_discrete_1} that $Q\geq
R-V'\bar{S}'\bar{\Pi}\bar{S}V = R+ V'\bar{L}V - V'\bar{\Pi}V $. 
Therefore \eq{Q\geq V'\bar{L}V\label{eq:split_discrete_water}} In view of this, to
complete the proof it is enough to show that $V'\bar{L}V$ is
positive definite.
This can be shown by the same  proof of Proposition \ref{prop:split_continuous_stable}.

% Since $\bar{L}$ is positive semi-definite, so is $V'\bar{L}V$. To
% show that $V'\bar{L}V$ is positive definite, let $z = \text{column}$ $
% \{z_1,z_2,\ldots, z_m\}$  be any vector such that $z'V'\bar{L}Vz=0$.
% Then $\bar{L}Vz = 0$. Since the kernel  of  $L$ is spanned
% $\mathbf{1}$, the kernel of $\bar{L}$   must  be spanned by
% $\mathbf{1}\otimes I_n$. It follows that
% $V_iz_i=V_jz_j,\;i,j\in\mathbf{m}$.  But because of joint
% observability, $\bigcap_{i\in\mathbf{m}} \mathcal{V}_i = 0$ so $V_iz_i =
% 0,\;i\in\mathbf{m}$. Thus $z_i = 0,\;i\in\mathbf{m}$
%  so $z=0$.
% Therefore   $V'\bar{L}V$ is positive definite.
Therefore $Q$ is
positive definite because of  \rep{eq:split_discrete_water}. From this and \rep{eq:split_discrete_1}
 it follows that \rep{eq:split_discrete_ly} is true.\hfill 
\qed

{ \bf Proof of Theorem~\ref{thm:split_discrete}:}
% Since the spectrum of each
% $\bar{A}_i+\bar{K}_i\bar{C}_i,\;i\in\mathbf{m}$, is assignable with
% $\bar{K}_i$, and $\widehat{A}_V(\tau)$  is a bounded matrix, to show
% that  for suitably defined $\bar{K}_i$ and  $q$ sufficiently large,
% the matrix $\Phi(\tau )$ defined in \rep{eq:split-discrete-phi}
%  converges to zero as fast as $\lambda^{\tau}$ does, it is sufficient to show that for $q$ sufficiently large,
%  $A_{V}(\tau )$ is a discrete-time stability matrix whose state-transition matrix converges to zero
% as fast as $\lambda^{\tau}$ does. {\color{red} It is of course obvious that the determinant of $\tilde A(V'\bar S(\tau)V)^q$ goes to zero as $q$ increases, but establishing a corresponding property for the transition matrix is more work. The intuition is that with a sufficiently large $q$, for any $\bar S(\tau)$, the spectral radius of $\bar A(V'\bar S(\tau)V)^q$ will be smaller, probably significantly smaller, than that of $\bar A$, and this spectral separation underpins the stability of the time-varying system associated with changing $\bar S(\tau)$. 
First it will be assumed that each $\bar{K}_i$ has been
selected so that  the the matrix
 $\bar{A}_V$ defined by \rep{eq:split_discrete_sunday}, is such that $\bar{A}_V^{\tau}$ converges
  to zero  as $\tau\rightarrow \infty$ as fast as $\lambda^{\tau}$ does.
 This can be done using standard
spectrum assignment techniques to  make the spectral radius of
$\bar{A}_V$ at least as small as $\lambda $.
 In view of   \rep{eq:split_discrete_as}, it is clear that to assign the convergence rate of the state transition matrix of
$\bar{A}(I_{mn}-P(I_{mn}-\bar{S}(\tau)))^q$ it is necessary and
sufficient to control the convergence rate
 of the state transition matrix of $A_{V}(\tau)$. This,  as we will now show, can be accomplished by choosing $q$ sufficiently large.
  We will actually detail two different ways to do this, each utilizing a different
matrix norm. Both approaches
 will be explained next
using the abbreviated notation  $B(\tau) = V'\bar{S}(\tau)V$; note
that with this simplification, $A_V(\tau) = \tilde{A}B^q(\tau)$
because of \rep{eq:split_discrete_ench}.

{\bf{Weighted Two-Norm:}}
For each fixed $\tau$ and each appropriately-sized matrix $M$, write
  $\|M\|_{R(\tau)}$ for the matrix norm  induced   by the vector norm
$\|x\|_{R(\tau)} \dfb \sqrt{x'R(\tau )x}$. Note that
$\|M\|_{R(\tau)}$ is the largest singular value of
 $R^{\frac{1}{2}}(\tau)MR^{-\frac{1}{2}}(\tau)$.
 Note in addition that
$$ (R^{\frac{1}{2}}(\tau)B(\tau)R^{-\frac{1}{2}}(\tau))'
  (R^{\frac{1}{2}}(\tau)B(\tau) R^{-\frac{1}{2}}(\tau)) < I
 $$
because of \rep{eq:split_discrete_ly}. This shows that the largest singular value of
$R^{\frac{1}{2}}(\tau)B(\tau) R^{-\frac{1}{2}}(\tau)$ is less than
one. Therefore \eq{\|B(\tau)\|_{R(\tau)}<1\label{ino}}

\vspace{-0.05in}
a) $\mathbb{N}$ is constant
\vspace{-0.1in}

In this case both $B(\tau)$  and $R(\tau)$ are constant, so it is
sufficient so choose
 choose $q$ so that
$\|\tilde{A}B^q(\tau)\|_{R(\tau)}\leq \lambda $. Since
$\|\cdot\|_{R(\tau )}$ is submultiplicative, this can be done by
choosing $q$ so that
\vspace{-0.1in}
\begin{equation}
\|B(\tau)\|^q_{R(\tau)} \leq \frac{\lambda}{\;\;\;\;\;\|\tilde{A}\|_{R(\tau)}}\label{eq:rnorm}    
\end{equation}
\vspace{-0.1in}
This can always be accomplished because of \rep{ino}.

\vspace{-0.05in}
b) $\mathbb{N}$ changes with time
\vspace{-0.1in}

In this case it is not possible to use the weighted two-norm
$\|\cdot\|_{R(\tau )}$ because it is time-dependent. A simple fix,
but perhaps not the most efficient one,  would be to use
 the standard two-norm $|\cdot|_2$ instead since it does not depend on time.
 Using this approach, the first step would be to
first choose, for each fixed $\tau$,
 an integer $p_1(\tau )
$ large enough so that $|B^{p_1(\tau)}(\tau )|_2<1$.  Such values of
$p_1(\tau)$  must exist because each $B(\tau)$ is
 a discrete-time stability
 matrix or equivalently, a matrix with a spectral radius less than $1$.
  Computing such a value amounts to looking at the
  largest singular value of $B^{p_1(\tau)}(\tau)$ for successively largest values of  $p_1(\tau)$
   until that singular value is less than $1$. Having accomplished this, a number $p$ can easily be computed
   so that $|B^p(\tau )|_2<1\;\forall \tau$ since there are only a finite number of
   distinct strongly connected graphs on $m$ vertices and consequently only a finite number of distinct
matrices   $B(\tau)$ in the set $\mathcal{B} = \{B(\tau ):\tau\geq 0\}$.
Choosing $p$ to be the maximum of the $p_1(\tau)$ with respect to $\tau
$ is thus a finite computation. The next step would be to compute
 an integer $\bar{p}$
 large enough so that each $|\tilde{A}(B^{p}(\tau))^{\bar{p}}|_2\leq \lambda$.
 A value of $q$ with the
 required property would then be $q = p\bar{p}$.

\vspace{-0.1in}
{\bf{Mixed Matrix Norm: }}
There is a different way to choose $q$ which   does
 not make use of either  Lemma \ref{lem:split_discrete_brian} or Proposition \ref{prop:split_discrete_mp}.  The approach exploits the
  ``mixed matrix norm'' introduced in \cite{ShaoshuaiTAC2015}. To define this norm requires several steps. \;
To begin, let $|\cdot|_{\infty}$
 denote   the standard induced
infinity  norm
    and write
 $\R^{mn\times mn}$ for  the vector space of all $m\times m$  block matrices $M = [M_{ij}]$
whose $ij$th entry     is a matrix $M_{ij}\in\R^{n\times n}$. With
$n_i = \dim \mathcal{V}_i,\;i \in\mathbf{m},$ and $\bar{n} =
n_1+n_2+\cdots n_m$,
 write $\R^{mn\times \bar{n}}$
for the vector space of all $m\times m$  block matrices $M =
[M_{ij}]$ whose $ij$th entry     is a matrix
$M_{ij}\in\R^{n\times n_j}$. Similarly write $\R^{ \bar{n}\times
mn}$ for the vector space of all $m\times m$  block matrices $M =
[M_{ij}]$ whose $ij$th entry     is a matrix
$M_{ij}\in\R^{n_i\times n}$. Finally write
 $\R^{\bar{n}\times \bar{n}}$ for  the vector space of all $m\times m$  block matrices $M = [M_{ij}]$
whose $ij$th entry     is a matrix $M_{ij}\in\R^{n_i\times n_j}$.

Note that $B\in \R^{mn\times mn}$, $\tilde{A}\in \R^{\bar{n}\times
\bar{n}}$, $V\in \R^{mn\times \bar{n}}$, and $V'\in\R^{\bar{n}\times
mn}$.
 For $M$ in any one of these four spaces, the {\em mixed matrix norm } \cite{ShaoshuaiTAC2015} of $M$, written $\|M\|$, is
 \vspace{-0.1in}
\eq{\|M\| = \|\langle M\rangle \|_{\infty}\label{eq:split_mmn}}
 where $\langle M\rangle $ is the  matrix in $\R^{m\times m}$  whose $ij$th entry is $\|M_{ij}\|_2$.
It is very easy to verify that $\|\cdot\|$ is in fact a norm. It is
even sub-multiplicative whenever matrix multiplication is defined.
Note in addition that $\|V\| = 1$ and $\|V'\| = 1$ because the
columns of each $V_i$ form an  orthonormal set.

\vspace{-0.1in}
Recall that $P = VV'$ is an orthogonal projection matrix. Using
this, the definition of $B(\tau)$ and the fact that $PV = V$, it is
easy to see that for any integer $p>0$,
$B^{p}(\tau) =V'(P\bar{S}(\tau)P)^pV$.
Thus
$ \|B^p(\tau)\|\leq \|(P\bar{S}(\tau)P)^{p}\|$.
Using this and  the  fact that the graph of $S'$ is strongly
connected,  one can conclude that for $p\geq (m-1)^2$,
$\|(P\bar{S}(\tau)P)^{p}\|<1.$
This is a direct consequence of  Proposition
 2 of \cite{ShaoshuaiTAC2015}. Thus
 \eq{\|B^p(\tau)\| <1 ,\;\;\;p\geq (m-1)^2\label{rajit}}

\vspace{-0.1in}
a) $\mathbb{N}$ is constant
\vspace{-0.1in}

In this case  $B(\tau)$  is constant so it is sufficient to choose
$q$ so that
 $\|\tilde{A}B^q(\tau)\|\leq \lambda $.
This can be done by choosing $q = p\bar{p}$  where $p\geq (m-1)^2$
and $\bar{p}$ is such that
 \eq{\|B^p(\tau)\|^{\bar{p}} \leq \frac{\lambda}{\|\tilde{A}\|}\label{rajit2}}
This can always be accomplished because of \rep{rajit}.

b) $\mathbb{N}$ changes with time

Note that \rep{rajit} holds for all $\tau $. Assuming $p$ is chosen
so that  $p\geq (m-1)^2$ it is thus possible to find, for each
$\tau$,  a positive integer $\bar{p}(\tau)$, for which
\eq{\|B^p(\tau)\|^{\bar{p}(\tau)} \leq
\frac{\lambda}{\|\tilde{A}\|}\label{rajit33}}

   Having accomplished this, a number $\bar{p}$ can easily be computed
   so that
\eq{\|B^p(\tau)\|^{\bar{p}} \leq
\frac{\lambda}{\|\tilde{A}\|}\label{rajit3}} holds for all $\tau$,
 since there  there are only a finite number of
   distinct strongly connected graphs on $m$ vertices and consequently only a finite number of distinct
matrices   $B(\tau)$ in the set $\mathcal{B} $ defined earlier.
 Choosing $\bar{p}$ to be the maximum of
$\bar{p}(\tau)$ with respect to $\tau $ is thus a finite
computation.
 A value of $q$ with the
 required property would then be $q = p\bar{p}$. \hfill $\qed$

\end{appendices}

\bibliographystyle{unsrt}
\bibliography{my,steve,observer2018}

\end{document}

%% file: steve.tex
\def\rep#1{(\ref{#1})}

\newcommand{\R}{\mathbb{R}}

\def\send#1#2{\stackrel{#1}{\hbox to #2{\rightarrowfill}}}
\def\-{\!\!\!\!\!-}

 \def\qed{ \rule{.08in}{.08in}}
\def\eq#1{\begin{equation}#1\end{equation}}

\newcommand{\dfb}{\stackrel{\Delta}{=}}

\newtheorem{theorem}{Theorem}
\newtheorem{lemma}{Lemma}
\newtheorem{remark}{Remark}
\newtheorem{proposition}{Proposition}

\def\qed{ \rule{.1in}{.1in}}

\def\R{{\rm I\!R}} 

\newcounter{seqn}[equation]
\def\theseqn{\arabic{equation}\alph{seqn}}

\def\endseqn{\eqno \@seqnnum
$$\ignorespaces}
\def\@seqnnum{(\theseqn)}

\newskip\mcentering \mcentering=0pt plus 1000pt minus 1000pt

\def\meqalignno#1{
\halign to\displaywidth{
    \hbox to 0pt{\kern\displaywidth\llap{$##$}\hss}\tabskip=\mcentering
    &\hfil$\displaystyle{##}$\tabskip=\mcentering
   &&$\displaystyle{{}##}$\hfil\tabskip=\mcentering
    \crcr
    #1\crcr}}

\def\rep#1{(\ref{#1})}

\def\eq#1{\begin{equation}#1\end{equation}}
\def\dspace{\multiply\normalbaselineskip 150
          \divide\normalbaselineskip 100 \normalbaselines
          \csname @@normalbaselineskip\endcsname\normalbaselineskip}
\def\sspace{\multiply\normalbaselineskip 200
         \divide\normalbaselineskip 300 \normalbaselines
         \csname @@normalbaselineskip\endcsname\normalbaselineskip}
\def\sdspace{\multiply\normalbaselineskip 160
         \divide\normalbaselineskip 150 \normalbaselines
         \csname @@normalbaselineskip\endcsname\normalbaselineskip}

\def\@{\tilde}

\def\3dot#1{\buildrel\textstyle...\over#1}